\documentclass[a4paper, 12pt, nofootinbib, amsmath, amssymb, aps, prd, onehalfspacing]{revtex4-2}

\usepackage{setspace}
\usepackage{CJKutf8} %Chinese characters
\usepackage{mathbbol} %mathbb \Omega
\usepackage{amssymb}

\usepackage[dvipsnames]{xcolor} % accessing additional colors
\usepackage{hyperref} %Automatically links \label and \ref commands; Always load last
\hypersetup{
    colorlinks=true,       % false: boxed links; true: colored links
    linkcolor=RoyalPurple,          % color of internal links
    citecolor=violet,        % color of links to bibliography
    filecolor=magenta,      % color of file links
    urlcolor=purple,           % color of external links
    anchorcolor=blue,
    menucolor=Fuchsia,
    runcolor=magenta
}
\usepackage[all]{hypcap} %Link navagates to top of figure instead of caption (below fig)

\usepackage{amsmath}
\usepackage{tikz}
\usepackage{nccmath}

\newcommand\numberthis{\addtocounter{equation}{1}\tag{\theequation}} % adding number tag to align*

\newcommand{\mycomment}[1]{}
\let\lc=\g

\newcommand{\na}{\nabla}

\newcommand{\be}{\begin{equation}}
\newcommand{\ee}{\end{equation}}
\newcommand{\bea}{\begin{eqnarray}}
\newcommand{\eea}{\end{eqnarray}}
\newcommand{\bear}{\begin{equation}\begin{aligned}}
\newcommand{\eear}{\end{aligned}\end{equation}}
\newcommand{\bes}{\begin{subequations}\begin{align}}
\newcommand{\ees}{\end{align}\end{subequations}}

\let\a=\alpha \let\b=\beta \let\g=\gamma \let\d=\delta
\let\z=\zeta     \let\l=\lambda
\let\m=\mu \let\n=\nu \let\x=\xi \let\r=\rho
\let\s=\sigma \let\t=\tau   
  
\let\G=\Gamma \let\D=\Delta

\newcommand{\polov}{\frac{1}{2}}

\newcommand{\dis}{\Phi}

\usepackage{comment}
\usepackage{setspace}

\usepackage{graphicx}% Include figure files
\usepackage{dcolumn}% Align table columns on decimal point
\usepackage{bm}% bold math

\def\ie{{\it i.e.\;}}

\def\Tr{{\rm Tr}}

\newcommand{\nn}{\nonumber}

\def\cL{{\cal L}}

\newcommand{\lc}{\Gamma}

\def\cD{{\cal D}}

\def\cL{{\cal L}}

\def\cO{{\cal O}}

\newcommand{\trp}[2]{\mathrm{tr}_{(#1#2)}\dis}
\newcommand{\divp}[1]{\mathrm{div}_{(#1)}\dis}
\newcommand{\divdivp}[2]{\mathrm{div}_{(#1#2)}\dis}
\newcommand{\divtrp}[2]{\mathrm{div}\,\mathrm{tr}_{(#1#2)}\dis}
\newcommand{\trdivp}[1]{\mathrm{tr}\mathrm{div}_{(#1)}\dis}

\newcommand{\trT}[2]{\mathrm{tr}_{(#1#2)}T}
\newcommand{\divT}[1]{\mathrm{div}_{(#1)}T}

\newcommand{\trdivT}[1]{\mathrm{tr}\mathrm{div}_{(#1)}T}

\newcommand{\DivT}[1]{\mathrm{Div}_{(#1)}T}

\newcommand{\trDivT}[1]{\mathrm{tr}\mathrm{Div}_{(#1)}T}

\newcommand{\trQ}[2]{\mathrm{tr}_{(#1#2)}Q}
\newcommand{\divQ}[1]{\mathrm{div}_{(#1)}Q}

\newcommand{\trdivQ}[1]{\mathrm{tr}\mathrm{div}_{(#1)}Q}

\newcommand{\DivQ}[1]{\mathrm{Div}_{(#1)}Q}

\newcommand{\trDivQ}[1]{\mathrm{tr}\mathrm{Div}_{(#1)}Q}

\begin{document}
\begin{CJK*}{UTF8}{gbsn} %Chinese characters

\title{On the renormalization of Metric-Affine Gravity theories}
\author{Oleg Melichev}
\email{melichev@shanghaitech.edu.cn}
\affiliation{\vspace{2mm}School of Physical Science and Technology, \\ ShanghaiTech University 上海科技大学, \\ 100 Haike Road, Pudong New Area, Shanghai, 201210, China\vspace{1mm}}
\begin{abstract}
We discuss the renormalization group in the context of gravitational theories with independent metric and affine connection.
Considering a class of theories with both propagating torsion and nonmetricity, we perform an explicit computation of one-loop divergences, starting from a simple yet phenomenologically viable modification of the Yang--Mills-like action.
Similarly to what happens in Poincar\'e gauge theory, in addition to the action, quadratic in curvature, torsion, and nonmetricity, many more terms are generated.
We correct a known result for the beta function of the Yang--Mills term and show that considerations previously presented in the literature are incomplete.
\end{abstract}

\maketitle

%\tableofcontents

%%%%%%%%%%%%%%%%%%%%%%%%%%%%%%%%%%%%%%%%%%%%%
\section{\label{sec:intro}Introduction}
%%%%%%%%%%%%%%%%%%%%%%%%%%%%%%%%%%%%%%%%%%%%%

Metric-Affine Gravity theories (MAGs) are a class of gravitational models where (certain components of) affine connection contain nontrivial dynamics in addition to the metric \cite{Utiyama:1956sy, Hehl:1976kj, Hehl:1994ue, Blagojevic:2012bc, Blagojevic:2013xpa, Einstein:1929b}.
In general, they may possess nonvanishing curvature, torsion, nonmetricity, or any combination thereof.
Many of them naturally recover General Relativity (GR) at low energies due to a particular version of the Higgs phenomenon \cite{Percacci:1990wy, Kirsch:2005st, Leclerc:2005qc, Percacci:2009ij}, while their high energy spectrum may contain several additional particles \cite{Percacci:2020ddy, Baldazzi:2021kaf,
Mikura:2023ruz,
Mikura:2024mji}.

If considered as a quantum theory, MAG faces the same issue of perturbative nonrenormalizability as metric gravity.
Starting from a single term linear in curvature, known as the Palatini action, 
the mass dimension of the perturbative coupling becomes negative, which leads to an infinite series of new terms generated by loop corrections.
Higher-order curvature terms can generate ghosts \cite{Percacci:2020ddy, Marzo:2021iok}.
This statement obviously holds in the first-order formalism.

On the other hand, MAG and its subclasses (such as Einstein--Cartan and Teleparallel theories) possess a rich phenomenological potential that cannot be simply ignored 
\cite{Benisty:2021sul, Jimenez-Cano:2022arz, Bahamonde:2021gfp, Hohmann:2018vle, Hohmann:2018dqh, Hohmann:2018ijr, Iosifidis:2019dua, Barrientos:2018cnx, Poplawski:2010kb, Poplawski:2011jz}.
The large number of independent invariants that one can construct at a given order leads to a large arbitrariness in the choice of the action.
When a specific form of classical action is chosen, one can study the dynamics and deviations from GR.
Certain versions of MAG do appear to be interesting candidates for resolving multiple cosmological and gravitational conundrums, such as black hole and Big Bang singularity problems \cite{Poplawski:2012ab, Cai:2015emx}, describing Dark Energy and/or Dark Matter \cite{Cai:2015emx}, as a natural candidate for inflaton \cite{Racioppi:2024zva},
and a possible solution to  the strong CP-problem \cite{Karananas:2024xja}.

It is of interest to study the quantum properties of such theories, starting from models that are relatively simple to work with.
The first question would be to see whether a given theory is perturbatively renormalizable, which can be understood using simple power counting arguments together with an explicit calculation of loop counterterms.
From the Effective Field Theory (EFT) point of view, nonrenormalizable theories are just as useful as renormalizable ones if the energy of the process at consideration is much lower than the energy of perturbative unitarity breakdown, \ie the EFT cut-off scale.
Different scenarios can be realized in MAG, depending on the mass hierarchy between this scale and the masses of the new particle states, produced by torsion and nonmetricity.
At energies below these masses, these states would already be “integrated out”, and quantum divergences give information about the behavior of Newton's coupling.
At energies comparable to the masses or above, loop divergences can modify the kinetic terms of those states.
If one has a theory with certain desirable properties, for example, free of ghosts, it is desirable to have such a description that keeps unitarity explicit at the quantum level, see \cite{Marzo:2021iok, Melichev:2023lpn} for earlier consideration of possible issues.

The landscape of MAGs is vast and hard to study even at the classical level.
The earliest systematic classification of Lagrangian contributions was given in \cite{Christensen:1979ue}, but later this research line was abandoned as too cumbersome.
A study that also included nonmetricity in the consideration was performed in \cite{Baldazzi:2021kaf}.
The terms are classified according to their mass dimension, which is expected to correspond to their ``relevance" (= the importance of their contributions to low-energy observables).
The mass dimension of the affine connection is one, which means that the dynamics of mass dimension two contributions is trivial, for neither torsion nor nonmetricity can propagate\footnote{This can change if, for example, when curvature is forced to vanish, in a way such as in Teleparallel theories. In this paper, we assume the absence of such kinematic restrictions.}. 
A systematic study of the renormalization group (RG) behavior of MAG with non-dynamical torsion and nonmetricity at the nonperturbative level was performed in \cite{Pagani:2015ema}.
The first nontrivial level corresponding to independent dynamics of the affine connection appears at order four in the mass dimension.
Such terms can be constructed with the Cartan curvature $F$ of the independent connection, the covariant derivative $D$ associated with it, and the tensorial difference between the independent connection and the Levi--Civita connection called distortion $\dis$.
The contributions can be schematically represented as follows:
\be \label{lag_dim4_schematic}
\cL = F_{....} F^{....} + D_. \dis_{...} D^. \dis^{...} + F_{...} D^. \dis^{...} 
+ F_{....} \dis_.{}^{..} \dis^{...}
+ \dis_{...} \dis_{..}{}^. D^.\dis^{...}
+ \dis_{...}\dis_{...}\dis^{...}\dis^{...} \, .
\ee
The dots represent Lorentz indices that can be contracted in any possible way, creating hundreds of terms, each of them coming with, in principle, an arbitrary coefficient.
Some of them can be rewritten in terms of others, leaving, however, 922 independent contributions \cite{Baldazzi:2021kaf}, in addition to the 12 contributions of mass dimension two and a possible cosmological constant.\footnote{This counting does not take into account the Schouten identity.}
The distortion is a linear function of the torsion and nonmetricity tensors, and one can think of these as all possible terms containing either.
Working with the fully general Lagrangians of the type described above is certainly complicated.
Due to the non-diagonal structure of the kinetic term, the RG flow coefficients (beta functions) appear as rational functions of the couplings in such a way that it is even hard to present results in a readable form.
It is of interest, however, to gain more intuition about quantum corrections coming from the independent components of the connection.
At this stage, we therefore need to restrict ourselves to a relatively simpler choice of the Lagrangian.
First considerations taking into account background with a nontrivial nonmetricity were presented in \cite{Donoghue:2016vck} with a simple Yang--Mills-like Lagrangian:
\be \label{lag_Donoghue}
S_{YM} = - \frac{c_1}{2} \int d^4 x \sqrt{g}\, F_{\m\n}{}^a{}_b F^{\m\n}{}_a{}^b \, ,
\ee
where a na\"ive calculation of the beta function was performed.
In this paper, we are going to look at a theory that includes \eqref{lag_Donoghue} as a subclass:\footnote{The reason for nonconsecutive numbering of the coefficients here and below is to keep an agreement with \cite{Percacci:2020ddy} and \cite{Baldazzi:2021kaf}.}
\be \label{lag_XYZ_ours}
S = - \frac{1}{2} \int d^4 x \sqrt{g} \left[ - a_0 F + c_1 F_{\m\n}{}^a{}_b F^{\m\n}{}_a{}^b 
+ a_1 T_\m{}^a{}_\n T^\m{}_a{}^\n 
+ a_4 Q_{\m a b} Q^{\m a b}
\right] \, .
\ee
Here $T$ and $Q$ denote torsion and nonmetricity tensors, indices are raised and lowered by means of the metric tensor, and $g = |\text{det}\, g_{\m\n}|$.
This Lagrangian is constructed in the spirit of Poincar\'e gauge theories, generalized to also accommodate the propagating nonmetricity \cite{Hayashi:1973mj, Cho:1976fr, Hehl:1978yt, Hayashi:1979wj, Pilch:1979bi}.
Our motivation for the inclusion of the dimension terms is twofold. 
First, they make torsion and nonmetricity massive and recover GR and low energies.
The second is a (very significant) technical convenience: in a special choice of variables and gauge, Lagrangian \eqref{lag_XYZ_ours} leads to a minimal second order Laplace type kinetic operator.\footnote{We call an operator minimal if all the derivatives entering its principal part are contracted with each other and nonminimal otherwise.}
This is a non-trivially pleasant feature, looking from the perspective of the full general MAG landscape.
For such operators the early proper time Seeley--DeWitt expansion is well studied, which allows for a relatively straightforward application of the heat kernel technique.
Hereafter we perform an explicit computation of the logarithmically divergent counterterms generated by the action \eqref{lag_XYZ_ours}, explaining why the previous computation presented in \cite{Donoghue:2016vck} was incorrect.
A similar computation for the Poincar\'e gauge theory with vanishing nonmetricity has been performed in \cite{Melichev:2023lwj}, and a computation of the running of the Starobinsky term $R^2$ in a more general class of theories with propagating torsion was performed in \cite{Martini:2023apm}.

Already looking at the schematic representation \eqref{lag_dim4_schematic} we can expect the appearance of many if not all such structures as loop corrections, even if we start from relatively simple Lagrangians containing few terms.
The calculations performed in the previous work, as well as during the preparation of this paper, confirm these expectations.
In order not to overwhelm ourselves with excessively long expressions, we will focus on terms that can modify the two-point function around flat space. 
They correspond to the first three terms in \eqref{lag_dim4_schematic}.
Other contributions of the types schematically presented by the last three terms, as well as all higher dimensional contributions, will be seen as less relevant and often neglected.

We employ the background field method and the generalized Schwinger--DeWitt technique \cite{Schwinger:1951nm, DeWitt:1964mxt, 
Barvinsky:1985an} based on the proper time method \cite{Fock:1937dy, 
Nambu:1950rs, Schwinger:1951nm} to compute the logarithmically divergent part of the effective action at one loop. 
Although it is perfectly possible to perform the computation of loop divergences in the usual coordinate basis, in order to be able to work with minimal kinetic operators exclusively, consideration of MAG as a gauge theory is advantageous for a certain type of Lagrangians, which includes the model under consideration.
We, therefore, give a brief introduction to the geometric picture of MAG in section \ref{sec:mag_as_gauge_theory}.
We proceed with a review of the theory at the level of Lagrangians
in section \ref{sec:general_Lag}.
We perform a perturbative expansion, introduce gauge fixing, and study the ghost sector in section \ref{sec:loop_calculation}.
We extract the results for the off-shell beta functions in section \ref{sec:beta_functions}.
We perform the on-shell reduction of the effective action in section \ref{sec:on_shell}, and finish with a discussion section \ref{sec:Discussion}.

%%%%%%%%%%%%%%%%%%%%%%%%%%%%%%%%%%%%%%%%%%%%%%
\section{\label{sec:mag_as_gauge_theory}MAG as Gauge theory}
%%%%%%%%%%%%%%%%%%%%%%%%%%%%%%%%%%%%%%%%%%%%%%

All fundamental interactions, such as the strong interactions described by a Yang--Mills field, are defined on vector bundles over a four-dimensional manifold \emph{M}.
A special characteristic of the gravitational field that distinguishes it from other interactions is that it belongs to a vector bundle with fibers in $R^4$ that is isomorphic to the tangent bundle \emph{TM}.
This automatically implies that we can define a metric on \emph{E} which is called fiber metric $g$, a linear connection in \emph{E}, $A_{\m}{}^{a}{}_{b}$, and a linear isomorphism between the vector bundle and the tangent bundle, which is called a soldering form or frame field $\theta^{a}{}_{\m}$, all as independent dynamical fields \cite{Percacci:1986ui, Percacci:1984bq, Komar:1985ws, Dabrowski:1986cf, Neeman:1987jgg, Siegel:1993xq, Afonso:2017bxr, Aoki:2019rvi}.
We adopt a distinction in the notation of indices so that the Greek ones always enumerate coordinates on the tangent bundle and the Latin ones -- on the vector bundle.
The gauge group in this formulation is $\mathbb{R}^{1, 3} \rtimes GL(4)$, while the metric belongs to the coset space $GL(4)/O(1, 3)$. 
The fiber connection can be seen as a Yang--Mills field that however lies in an adjoint representation of a noncompact group $GL(4)$.

Alternatively, one can define the soldering using arbitrary bases $\{e_a\}$
in the tangent spaces
and $\{e^a\}$ in the cotangent spaces.
Given a coordinate system $x^\m$, 
they are related to the coordinate bases by
\be
e_a=\theta_a{}^\m\partial_\m \, , \quad
e^a=dx^\m \theta^{-1}{}_{\m}{}^{a} \, . 
\label{frames}
\ee
Then, we can construct a metric and connection on the tangent bundle \emph{TM} as:
\begin{subequations}
\begin{align} \label{pullbackmetric}
g_{\m\n} &= \theta^a{}_\m\, \theta^b{}_\n\, g_{ab} \, , 
\\ \label{pullbackconnection}
A_\lambda{}^\m{}_\n &= \theta_a{}^\m A_\lambda{}^a{}_b \theta^b{}_\n
+\theta_a{}^\m \partial_\lambda \theta^a{}_\n \, .
\end{align} 
\end{subequations}
Hereafter we abbreviate the inverse of the soldering (coframe field) as 
\be
\theta_b{}^\m \equiv \theta^{-1}{}^\m{}_b = g_{ab} \theta^a{}_{\n} g^{\n\m} \, . 
\ee
The action of the covariant derivative associated with the independent connection on a tensor of rank $(n,m)$ is
\be \label{covd_notation_acting} 
D_\m T^{\a_1 \dots \a_n}_{\b_1 \dots \b_m} = \partial_\m T^{\a_1 \dots \a_n}_{\b_1 \dots \b_m} + \sum_{i=1}^n A_\m{}^{\a_i}{}_\s T^{\a_1 \dots \s \dots \a_n}_{\b_1 \dots \b_m} - \sum_{j=1}^m A_\m{}^\s{}_{\b_j} T^{\a_1 \dots \a_n}_{\b_1 \dots \s \dots \b_m} \, .
\ee
An analogous formula holds for the vector bundle indices.
Using \eqref{pullbackmetric} we also observe that the determinant of metric that enters the action functionals \eqref{lag_Donoghue} and \eqref{lag_XYZ_ours} is
\be \label{detmetric_explanation}
\sqrt{g} = \sqrt{|\text{det}\, g_{\m\n}|} = \text{det}\, \theta^a{}_\m \sqrt{|\text{det}\, g_{ab}|} \, .
\ee
We define the Cartan curvature of the vector bundle as:
\be
\label{curvature_fiber_def}
F_{\m\n}{}^a{}_b=
\partial_\m A_\n{}^a{}_b
-\partial_\n A_\m{}^a{}_b
+A_\m{}^a{}_c A_\n{}^c{}_b
-A_\n{}^a{}_c A_\m{}^c{}_b \, , 
\ee
which represents the strength of the connection field.
Then, the standard Yang--Mills action will have the form \eqref{lag_Donoghue}.
In addition to that term, however, 
one can get 15 more such contributions by contracting the indices in different ways.
Furthermore, owing to the existence of the soldering, the term linear in curvature is also allowed:
\be
\label{action_Palatini}
S = \int d^4 x\; \sqrt{g}\, F_{\m\n}{}^a{}_b \theta_a{}^\m \theta^b{}_\r g^{\r\n} \, . 
\ee
It is called the Palatini action.
The existence of such a term is the distinctive feature of gravity.

Alongside curvature, there are two other independent tensorial characteristics of the manifold.
Torsion is defined as the exterior covariant derivative of the soldering
\be
\label{torsion_def}
T_\m{}^a{}_\n=
\partial_\m \theta^a{}_\n-\partial_\n \theta^a{}_\m+
A_\m{}^a{}_b\, \theta^b{}_\n-A_\n{}^a{}_b\, 
\theta^b{}_\m \, ,
\ee
and can be seen as the soldering field strength; while the nonmetricity is defined as the (minus) covariant derivative of the fiber metric
\be
\label{nonmetricity_def}
Q_{\lambda ab}=-D_\lambda g_{ab}= 
-\partial_\lambda g_{ab}
+A_\lambda{}^c{}_a\, g_{cb}
+A_\lambda{}^c{}_b\, g_{ac} \, , 
\ee
and can be seen as the metric field strength.
Given the soldering, any tensorial index can be moved from the vector bundle to the tangent bundle and vice versa.
For example, 
\be
F_{\m\n}{}^\a{}_\b = F_{\m\n}{}^a{}_b \theta_a{}^\a \theta^b{}_\b 
\ee
is the Cartan curvature of the tangent bundle.

Let us now come back to the discussion of the gauge group.
An arbitrary non-degenerate $\Lambda^a{}_b(x)$ 
describes a local change of frames $e'_a(x)=e_b(x)\Lambda^a{}_b(x)$, which is independent of the diffeomorphisms $x'(x)$.
The action of these transformations on the fields is given by
\bear
\label{transform_GL4_and_diffeos}
\theta^a{}_\m(x)\mapsto &{\theta^\prime}^a{}_\m(x^\prime)= 
\Lambda^{-1 a}{}_b(x)\, \theta^b{}_\n(x)
{\partial x^\n\over\partial x^{\prime \m}} \, ,
\\
g_{ab}(x)\mapsto &{g^\prime}_{ab}(x^\prime)=
\Lambda^c{}_a(x)\, \Lambda^d{}_b(x)\, g_{cd}(x) \, ,
\\
A_\m{}^a{}_b(x)\mapsto & A^\prime_\m{}^a{}_b(x^\prime)=
{\partial x^\n\over\partial x^{\prime \m}}
\left[\Lambda^{-1 a}{}_c(x) A_\n{}^c{}_d(x) \Lambda^d{}_b(x)+
\Lambda^{-1 a}{}_c(x)\partial_\n\Lambda^c{}_b(x)\right] \, . 
\eear
There are two common ways to partially fix this gauge:
\begin{itemize}
\item In \emph{metric gauge}, we demand
\be
\label{metric_gauge}
\theta_a{}^\m=\d_a^\m \, . 
\ee
After that one can just stop making any distinction between tangent and vector bundle indices.
The $GL(4)$ invariance is broken completely, and one is left with just diffeomorphisms.
This is the way gravity is normally described, in the absence of fermionic matter.
Torsion then becomes a purely algebraic object:
\be
T_\m{}^\r{}_\n=
A_\m{}^\r{}_\n-A_\n{}^\r{}_\m
\label{TA} \, ,
\ee
while the nonmetricity still involves a derivative of $g$.
\item In \emph{vierbein gauge} we demand instead
\be
\label{vierbein_gauge}
g_{ab}=\eta_{ab} \, . 
\ee
This breaks the gauge group to $\mathbb{R}^{1, 3} \rtimes O(1,3)$.
Then (\ref{pullbackmetric}) becomes the defining relation
for the tetrad (vierbein) and the connection in this case is called the ``spin connection''
\footnote{We stick to the convention that the components of
the same geometrical object in different bases
should not be given different names.}.
In this gauge, the nonmetricity is a purely algebraic object:
\be
Q_{cab}=A_{cab}+A_{cba}
\label{QA} \, ,
\ee
whereas torsion still involves a derivative of $\theta$.
\end{itemize}
In the following, we will also adopt more elaborate gauges.
However, in the majority of this paper, we will stick to the metric gauge, which will be implicitly assumed when formulae are written with Greek indices only, unless otherwise stated.

All said so far applies equally to MAG and metric gravity theories.
Whether one chooses to work with soldering, tetrads or metric is a gauge choice and does not influence the spectrum of the theory.
In addition to that, one has the freedom to impose kinematic constraints that reduce the configuration space.
In GR one \emph{a priori} assumes that torsion and nonmetricity vanish everywhere:
\be
\label{condition_no-torsion_no-nonmetricity}
Q_{\m\n\r} = T_{\m\n\r} = 0 \, . 
\ee
Such constraints are unnecessary from the conceptual point of view, and they can be enforced automatically as a consequence of mass suppression of the additional degrees of freedom.
This happens everywhere except a measure zero subset of the theory space \cite{Percacci:2009ij, Baldazzi:2021kaf, Melichev:2023lpn, Percacci:2023rbo}.
This implies that the correct dynamics describing gravity at high energies may differ from the standard spin-2 dynamics of GR and may involve independent connection degrees of freedom.

Given a vector bundle and frame field, 
there is a unique connection, 
called the Levi--Civita (LC) connection, which is torsionless and metric-compatible.
Its components are
\be \label{lc_connection_def}
\lc_{abc}=\frac12\left(
E_{acb}+E_{cab}-E_{bac}\right)
-\frac12\left(f_{abc}+f_{cab}-f_{bca}\right) \, ,
\ee
where 
\bear
E_{cab} &={\theta}_c{}^\lambda\, \partial_\lambda g_{ab} \, ,
\\
f_{bc}{}^a &=
\left(\theta_b{}^\m\,
\partial_\m\theta_c{}^\lambda-
\theta_c{}^\m\, \partial_\m\theta_b{}^\lambda
\right)\theta^a{}_\lambda \, . 
\eear
The covariant derivative associated with the LC connection will be denoted as $\nabla$.
One might prefer working with one connection or another depending on the context.

The independent affine connection can be uniquely decomposed into
\be
A_\m{}^a{}_b=\lc_\m{}^a{}_b+\dis_\m{}^a{}_b
\label{distorion_decomposition}
\ee
where $\lc_\m{}^a{}_b$ is the LC connection
and $\dis_\m{}^a{}_b$ is a tensor called the distortion.
In general, it does not have any symmetry properties.
From (\ref{torsion_def}) and (\ref{nonmetricity_def}) one finds
\be
\label{TQphi}
T_{\a\b\g} = \dis_{\a\b\g}-\dis_{\g\b\a} \, ,\qquad
Q_{\a\b\g} = \dis_{\a\b\g}+\dis_{\a\g\b} \, . 
\ee
These relations can be inverted, to give the distortion
as a function of the torsion and the nonmetricity.
In fact, we can write
\be
\label{discon}
\dis_{\a\b\g}=K_{\a\b\g} + L_{\a\b\g} \, ,
\ee
where $K$ and $L$ are the contortion and the disformation tensors respectively
\bea
K_{\a\b\g} &=& \frac{1}{2}\left(T_{\a\b\g}+T_{\b\a\g}-T_{\a\g\b}\right) \, , 
\\
L_{\a\b\g} &=& \frac{1}{2}\left(Q_{\a\b\g}+Q_{\g\b\a}-Q_{\b\a\g}\right) \, .
\eea

One can see from (\ref{pullbackconnection}) that the total covariant derivative of the soldering using the same connection on both indices is zero, a statement often called the ``tetrad postulate''.
In the following, we will use the ``total covariant derivative'' of the soldering defined using the LC connection for the Greek index and the independent dynamical connection for the Latin one, which is nothing but the distortion tensor:
\be \label{totcovd_of_soldering}
\cD_\m\theta^a{}_\r\equiv
\partial_\m\theta^a{}_\r-\lc_\m{}^\sigma{}_\r\theta^a{}_\sigma
+A_\m{}^a{}_b\theta^b{}_\r=\dis_\m{}^a{}_\r \, . 
\ee

The curvature of the LC connection
\be
\label{defR}
R_{\m\n}{}^a{}_b=
\partial_\m \lc_\n{}^a{}_b
-\partial_\n \lc_\m{}^a{}_b
+\lc_\m{}^a{}_c \lc_\n{}^c{}_b
-\lc_\n{}^a{}_c \lc_\m{}^c{}_b 
\ee
is the Riemann tensor and its components are related to the curvature tensor 
of the independent connection as follows:
\bea \label{FtoR}
F_{\m\n}{}^\a{}_\b & = & 
R_{\m\n}{}^\a{}_\b
+\nabla_{\m} \dis_{\n \,\,\, \b}^{\,\,\, \a}
-\nabla_{\n} \dis_{\m\,\,\, \b}^{\,\,\, \a}
+ \dis_{\m\,\,\, \g}^{\,\,\, \a} \dis_{\n \,\,\, \b}^{\,\,\, \g}
- \dis_{\n \,\,\, \g}^{\,\,\, \a} \dis_{\m \,\,\, \b}^{\,\,\, \g} \, ,
\eea
which is sometimes known as the post-Riemannian decomposition.
Note that our notations regarding the position of indices are somewhat different from the one that is common in gravitational literature, but the components of the Riemann curvature on the tangent bundle
$R_{\m\n}{}^\r{}_\l = \theta_a{}^\r R_{\m\n}{}^a{}_b \theta^b{}_\l$ agree with them due to its symmetry properties.
The other important property of the curvature tensors is that they satisfy the Bianchi identities (see \cite{McCrea:1992wa} for an extensive analysis):
\begin{subequations}
\begin{align}
F_{[\alpha\beta}{}^\gamma{}_{\delta]}
-D_{[\alpha} T_\beta{}^\gamma{}_{\delta]}
-T_{[\alpha}{}^\epsilon{}_{\beta|} T_\epsilon{}^\gamma{}_{|\delta]}&=0 \, ,
\label{bianchi1F}
\\
D_{[\alpha}F_{\beta\gamma]}{}^\delta{}_\epsilon+T_{[\alpha}{}^\eta{}_{\beta|} F_{\eta|\gamma]}{}^\delta{}_\epsilon&=0 \, ,
\label{bianchi2F}
\\
R_{[\alpha\beta}{}^\gamma{}_{\delta]}
&=0 \, ,
\label{bianchi1R}
\\
\nabla_{[\alpha}R_{\beta\gamma]}{}^\delta{}_\epsilon &=0 \, .
\label{bianchi2R}
\end{align}
\end{subequations}

Let us briefly discuss other tensorial structures that can be obtained from \eqref{curvature_fiber_def}, \eqref{torsion_def}, \eqref{nonmetricity_def}.
There are three possible ways of contracting indices of the Cartan curvature:
\begin{subequations}
\begin{align}
&F^{(13)}_{\,\n\b} = F_{\m\n}{}^\m{}_\b \, ,
\\
&F^{(34)}_{\,\m\n} = - F^{(34)}_{\,\n\m} = F_{\m\n}{}^\a{}_\a \, ,
\\
&F^{(14)}{}_\n{}^\a = g^{\m\b} F_{\m\n}{}^\a{}_\b{} \, ,
\end{align}
\end{subequations}
while the scalar contraction with the metric is unique:
\bear \label{FscalarToR}
F = g^{\n\b} F_{\m\n}{}^\m{}_{\b} & = & R
+ \nabla_\m \dis_\n{}^{\m\n} - \nabla_\n \dis_\m{}^{\m\n} +\dis_{\m \,\,\,\, \g}^{\,\,\,\, \m}\dis_{\n}^{\,\,\,\, \g\n}-\dis_{\n\m\g}\dis^{\m\g\n} \, .
\eear
Here we expressed the Cartan scalar via the Ricci scalar using \eqref{FtoR}.
The same expression can also be written in terms of torsion and nonmetricity 
\bear
\label{FtoRTQ}
F & = R
+\frac{1}{4}T_{\a\b\g}T^{\a\b\g}
+\frac{1}{2}T_{\a\b\g}T^{\a\g\b}
-T_\a T^\a
\\
 & 
+\frac{1}{4}Q_{\a\b\g}Q^{\a\b\g}-\frac{1}{2}Q_{\a\b\g}Q^{\b\a\g}
-\frac{1}{4} Q_{\a} Q^{\a}
+\frac{1}{2} \hat{Q}_{\a} Q^{\a} \\
& 
- Q_{\a\b\g}T^{\a\b\g}
 + Q_{\a} T^{\a}
 - \hat{Q}_{\a} T^{\a} + \text{total derivative} \, . 
\eear
where
\begin{subequations}
\begin{align}
\label{TQcontractions_def}
T_\a &= \trT23_\a = - \trT12_\a = T_\a{}^\r{}_\r \, , \\
Q_\m &= \trQ23_\m = Q_\m{}^\a{}_\a \, , \\
\hat{Q}_\m &= \trQ12_\m = Q^\a{}_{\a\m} \, .
\end{align}
\end{subequations}
The following table summarizes our notation,
which is Yang--Mills-like for the independent connection,
and GR-like for the LC connection:
\begin{center}
\begin{tabular}{|c|c|c|c|}
\hline
& coefficients 
& cov.\! der.
& curvature 
\\
\hline
Levi--Civita connection 
& $\lc_\m{}^\r{}_\sigma$
& $\nabla_\m$ 
& $R_{\m\n}{}^\r{}_\sigma$
\\
\hline
Independent dynamical connection 
& $A_\m{}^\r{}_\sigma$
& $D_\m$ 
& $F_{\m\n}{}^\r{}_\sigma$
\\
\hline
\end{tabular}
\end{center}
We stress that the same notation is used
when the connection coefficients refer to a coordinate basis (tangent bundle, Greek indices) or an orthonormal basis (vector bundle, Latin indices).
In the next section, we discuss construction of the action, which we assume to be local Lorentz scalar and polynomial in tensorial fields and field strengths ($F$, $R$, $T$, $Q$, $\dis$).

%%%%%%%%%%%%%%%%%%%%%%%%%%%%%%%
\section{\label{sec:general_Lag}Lagrangians}
%%%%%%%%%%%%%%%%%%%%%%%%%%%%%%%

We follow notations of \cite{Baldazzi:2021kaf} and list independent invariants, starting with mass dimension two, at which we have 11 terms:
\bear
M^{TT}_1&=T^{\m\r\n}T_{\m\r\n} \, , \quad
M^{TT}_2= T^{\m\r\n}T_{\m\n\r} \, , \quad
M^{TT}_3= \trT12^{\m} \trT12_\m \, , 
 \\
M^{QQ}_1&=Q^{\r\m\n}Q_{\r\m\n} \, , \quad 
M^{QQ}_2=Q^{\r\m\n}Q_{\n\m\r} \, , 
\\
M^{QQ}_3&=\trQ23^{\m} \trQ23_\m \, , \quad
M^{QQ}_4=\trQ12^{\m}\trQ12_\m \, , \quad
M^{QQ}_5=\trQ23^{\m}\trQ12_\m \, , 
\\
M^{TQ}_1&=T^{\m\r\n} Q_{\m\r\n} \, , \quad 
M^{TQ}_2=\trT12^{\m}\trQ23_\m \, , \quad
M^{TQ}_3=\trT12^{\m}\trQ12_\m \, . 
\label{ttqq}
\eear
Then, the lowest-order Lagrangian (apart from the cosmological constant) can be written as
\be \label{Lagrangian_dim2_Cartan}
\cL^{(\text{dim}\, 2)}_C \left[ g, A \right] =-\polov\left[-m^F F+\sum_{i=1}^{3}a^{TT}_i M^{TT}_i
+\sum_{i=1}^{3}a^{TQ}_i M^{TQ}_i
+\sum_{i=1}^{5}a^{QQ}_i M^{QQ}_i\right] \, .
\ee
The first term is the Palatini one.
Performing its post-Riemannian expansion \eqref{FtoRTQ} we can write the same Lagrangian as:
\be \label{Lagrangian_dim2_EinsteinRTQ}
\cL^{(\text{dim}\, 2)}_E \left[ g, T, Q \right] =-\polov\left[-m^R R+\sum_{i=1}^{3}m^{TT}_i M^{TT}_i
+\sum_{i=1}^{3}m^{TQ}_i M^{TQ}_i
+\sum_{i=1}^{5}m^{QQ}_i M^{QQ}_i\right] \, . 
\ee
The correspondence between the parameters $m_i$ and $a_i$ is
\bear
&m^R = m^F \, , \quad m^{TT}_1 = a^{TT}_1 - \frac{m^F}{4} \, , \quad m^{TT}_2 = a^{TT}_2 - \frac{m^F}{2} \, , \quad m^{TT}_3 = a^{TT}_3 - m^F \, , 
\\
&m^{TQ}_1 = a^{TQ}_1 + m^F \, , \quad m^{TQ}_2 = a^{TQ}_2 + m^F \, , \quad m^{TQ}_3 = a^{TQ}_3 - m^F \, , 
\quad m^{QQ}_1 = a^{QQ}_1 - \frac{m^F}{4} \, ,
\\
&m^{QQ}_2 = a^{QQ}_2 + \frac{m^F}{2} \, ,
\quad m^{QQ}_3 = a^{QQ}_3 + \frac{m^F}{4} \, ,
\quad m^{QQ}_4 = a^{QQ}_4 \, ,
\quad m^{QQ}_5 = a^{QQ}_5 - \frac{m^F}{2} \, .
\eear
The two Lagrangians \eqref{Lagrangian_dim2_Cartan} and \eqref{Lagrangian_dim2_EinsteinRTQ} describe the same physics, but they are different in their forms.
The former one is written using Cartan curvature and therefore is a functional of the metric and the independent affine connection.
Torsion and nonmetricity tensors are then functions of the connection \eqref{TQphi}.
The latter one, instead, contains the Ricci scalar which is a function of metric only.
In this form, torsion and nonmetricity are independent variables, that can be seen as a complicated form of matter.

In what follows, we will say that expressions containing the Levi--Civita connection, Riemann curvatures, and either distortion $\dis$ or a pair $(T, Q)$ are in the \textbf{Einstein form}, while expressions that treat the affine connection as an independent dynamical variable will be said to be in the \textbf{Cartan form}.

For the sake of clarity, we use the following abbreviations for tensorial contractions.
Given any tensor $\dis_{abc}$, we define
\bear
&\trp12_c\equiv\phi^{(12)}_c=\dis_a{}^a{}_c \, , 
\qquad 
\trp13^b\equiv\dis^{(13)b}=\dis_a{}^{ba}\, , \mathrm{etc.} , \\
&\divp1^b{}_c=\nabla_a\dis^{ab}{}_{c}\, , 
\qquad
\divp2_{ac}=\nabla_b\dis_a{}^b{}_c\, , \ \mathrm{etc.} ,
\\
&\divdivp23_c=\nabla_a\nabla_b\dis^{ab}{}_c\, , \ \mathrm{etc.} 
\qquad
\trdivp1=\divp1^a{}_a\, , 
\\
&\divtrp12=\nabla_a\trp12^a\, , \ \mathrm{etc.} \nonumber
\eear
Note that with the LC connection
$\divtrp12=\trdivp3$, etc.
We will write ``Div" instead of ``div" when we mean the same expression but with the independent connection, although this difference will not affect any of the formulae presented in the paper.

In the Cartan form, we have 16 terms of the $FF$-type:
\bear
\label{LFF}
L^{FF}_1&=F^{\mu\nu\rho\sigma}F_{\mu\nu\rho\sigma}\, , \quad
L^{FF}_2=F^{\mu\nu\rho\sigma}F_{\mu\nu\sigma\rho}\, , \quad
L^{FF}_3=F^{\mu\nu\rho\sigma}F_{\rho\sigma\mu\nu} \, , 
\\
L^{FF}_4&=F^{\mu\nu\rho\sigma}F_{\mu\rho\nu\sigma}\, , \quad
L^{FF}_5=F^{\mu\nu\rho\sigma}F_{\mu\sigma\nu\rho}\, , \quad
L^{FF}_6=F^{\mu\nu\rho\sigma}F_{\mu\sigma\rho\nu} \, , 
\\
L^{FF}_7&=F^{(13)\mu\nu}F^{(13)}_{\mu\nu}\, , \quad
L^{FF}_8=F^{(13)\mu\nu}F^{(13)}_{\nu\mu}\, , 
\\
L^{FF}_9&=F^{(14)\mu\nu}F^{(14)}_{\mu\nu}\, , \quad 
L^{FF}_{10}=F^{(14)\mu\nu}F^{(14)}_{\nu\mu}\, , 
\\ 
L^{FF}_{11}&=F^{(13)\mu\nu}F^{(14)}_{\mu\nu}\, , \quad
L^{FF}_{12}=F^{(13)\mu\nu}F^{(14)}_{\nu\mu}\, , 
\\
L^{FF}_{13}&=F^{(34)\mu\nu}F^{(34)}_{\mu\nu}\, , \quad
L^{FF}_{14}=F^{(34)\mu\nu}F^{(13)}_{\mu\nu}\, , \quad
L^{FF}_{15}=F^{(34)\mu\nu}F^{(14)}_{\mu\nu}\, , 
\\
L^{FF}_{16}&=F^2\ ;
\eear
9 terms of the $FDT$-type:
\be
\label{LFT}
\begin{tabular}{lll}
$L^{FT}_1=F^{\mu\nu\rho\sigma}D_\mu T_{\nu\rho\sigma}$ ,
&
$L^{FT}_{10}=F^{(14)\mu\nu} D_\mu\trT12_{\nu}$ , 
& $L^{FT}_{11}=F^{(14)\mu\nu} D_\nu\trT12_{\mu}$ , 
\\
$L^{FT}_{12}=F^{(34)\mu\nu} D_\mu\trT12_{\nu}$ , 
&
$L^{FT}_{13}=F^{(13)\mu\nu}\, \DivT1_{\mu\nu}$ , 
& $L^{FT}_{14}=F^{(13)\mu\nu}\, \DivT1_{\nu\mu}$ , 
\\
$L^{FT}_{15}=F^{(14)\mu\nu}\, \DivT1_{\mu\nu}$ , 
& $L^{FT}_{17}=F^{(13)\mu\nu}\, \DivT2_{\mu\nu}$ , 
& $L^{FT}_{21}=F\, \trDivT1$ ;
\end{tabular}
\ee
9 terms of the $FDQ$-type:
\be
\begin{tabular}{llll}
& $L^{FQ}_{10}=F^{(14)\mu\nu} D_\mu\trQ12_{\nu}$ , 
& $L^{FQ}_{11}=F^{(14)\mu\nu} D_\nu\trQ12_{\mu}$ , 
& $L^{FQ}_{12}=F^{(14)\mu\nu} D_\mu\trQ23_{\nu}$ , \\
& $L^{FQ}_{14}=F^{(34)\mu\nu} D_\mu\trQ12_{\nu}$ , 
& $L^{FQ}_{16}=F^{(13)\mu\nu}\, \DivQ1_{\mu\nu}$ , 
& $L^{FQ}_{17}=F^{(14)\mu\nu}\, \DivQ1_{\mu\nu}$ , \\
& $L^{FQ}_{18}=F^{(13)\mu\nu}\, \DivQ2_{\mu\nu}$ , 
& $L^{FQ}_{19}=F^{(13)\mu\nu}\, \DivQ2_{\nu\mu}$ , 
& $L^{FQ}_{23}=F\, \trDivQ1$ ;
\end{tabular}
\label{LFQ}
\ee
4 terms of the $DTDT$-type:
\bear
&L^{TT}_{1} = D^\a T^{\b\g\d} D_\a T_{\b\g\d} \, ,
\quad\quad\quad\;\;\;
L^{TT}_{2} = D^\a T^{\b\g\d} D_\a T_{\b\d\g} \, ,
\\
&L^{TT}_{3} = D^\a \trT12^\b D_\a\trT12_\b \, , \,
\quad L^{TT}_{5} = \DivT1^{\a\b}\DivT1_{\b\a} \, ; 
\label{LTT}
\eear
5 terms of the $(DQ)^2$-type:
\be
\begin{tabular}{lll}
$L^{QQ}_{1} = D^\alpha Q^{\beta\gamma\delta}\, 
D_\alpha Q_{\beta\gamma\delta}$ , 
&
$L^{QQ}_{10} = %\DivDivQ12^\alpha\, \trQ12_\alpha$
\DivQ2^{\alpha\beta}\, D_\alpha\trQ12_\beta$ , 
\\
$L^{QQ}_{11} = %\DivDivQ12^\alpha\, \trQ23_\alpha$
\DivQ2^{\alpha\beta}\, D_\alpha\trQ23_\beta$ , \quad
&
$L^{QQ}_{12} = %\DivDivQ23^\alpha\, \trQ12_\alpha$
\DivQ2^{\alpha\beta}\, D_\beta\trQ12_\alpha$ ,\quad 
& \quad
$L^{QQ}_{14} = (\trDivQ1)^2$ ;
\end{tabular}
\label{LQQ}
\ee
and 4 terms of the $DTDQ$-type:
\be
\begin{tabular}{lll}
$L^{TQ}_{1} = D^\alpha T^{\beta\gamma\delta}\, 
D_\alpha Q_{\beta\gamma\delta}$ , 
&\quad
$L^{TQ}_{10} = %\trT12_\alpha\, \DivDivQ12^\alpha$
\DivQ2^{\alpha\beta} D_\alpha\trT12_\beta$ , 
\\
$L^{TQ}_{11} = %\trT12_\alpha\, \DivDivQ23^\alpha$
\DivQ2^{\alpha\beta} D_\beta\trT12_\alpha$ , 
&\quad
$L^{TQ}_{12} = %D_\beta \trT12^\beta\, D_\alpha\trQ12^\alpha$
\trDivT1\, \trDivQ1$ .
\end{tabular}
\label{LTQ}
\ee
The non-consecutive numbering is due to the way we defined bases for general MAGs in \cite{Baldazzi:2021kaf}.
In fact, one can construct many more invariants of the same mass dimension.
However, all other invariants can be expressed via those that we have listed and, in some cases, some other invariants that do not contribute to the two-point function around flat space.
For instance, some of the terms of $FDT$ and $FDQ$ types can be reduced using the Bianchi identities \eqref{bianchi1F}, \eqref{bianchi2F}; while other terms such as $DTDT$ can be reexpressed using the decomposition \eqref{FtoR}.
Such relations have been thoroughly listed in \cite{Baldazzi:2021kaf}.
Here we work in the basis listed above and refrain from listing such additional invariants that can be rewritten in terms of others.

Putting all together, the dimension-four contribution in Cartan form in the chosen basis is then
\bear \label{Lag_dim4_Cartan}
\cL^{(\text{dim}\, 4)}_C \left[ g, A \right] &=
\frac12 \Big[
\sum_{i=1}^{16} c^{FF}_i L^{FF}_i
+
\sum_{i=1,10-15,17,21} c^{FT}_i L^{FT}_i
+\sum_{i=10-12,14,16-19,23} c^{FQ}_i L^{FQ}_i
\\ &
+\sum_{i=1,2,3,5} c^{TT}_i L^{TT}_i
+\sum_{i=1,10,11,12} c^{TQ}_i L^{TQ}_i
+\sum_{i=1,10,11,12,14} c^{QQ}_i L^{QQ}_i
+ \dots
\Big] \, .
\eear
The dots represent terms such as $FTT$, $TTDT$, $QQQQ$, etc. that do not contribute to the flat space two-point function.

The Cartan form offers a clear geometrical interpretation.
On the other hand, it is often desirable not to use any structures related to the connection that is not compatible with metric, simply because the chances of making a mistake increase otherwise.
For this reason, the Einstein form is preferable in the intermediate steps of calculations.
The contractions we use are similar to the ones in Cartan form, but there is a smaller number of invariants.
In the kinetic torsion sector, we have nine $(\nabla T)^2$ terms
\be
\begin{tabular}{lll}
\label{Lag_terms_HTT}
$H^{TT}_{1} = \nabla^\a T^{\b\g\d} \nabla_\a T_{\b\g\d}$ ,
& $H^{TT}_{2} = \nabla^\a T^{\b\g\d} \nabla_\a T_{\b\d\g}$ ,
\\
$H^{TT}_{3} = \nabla^\a\trT12^\b \nabla_\a\trT12_\b$ ,
\\
$H^{TT}_{4} = \divT1^{\a\b}\divT1_{\a\b}$ ,
& $H^{TT}_{5} = \divT1^{\a\b}\divT1_{\b\a}$ ,
\\
$H^{TT}_{6} = \divT2^{\a\b}\divT2_{\a\b}$ ,
& $H^{TT}_{7} = \divT1^{\a\b}\divT2_{\a\b}$ ,
\\
$H^{TT}_{8} = %- \divdivT12^\a\trT12_\a$
\divT2^{\a\b}\nabla_\a\trT12_\b$ ,
& $H^{TT}_{9} = (\trdivT1)^2$ ,
\end{tabular}
\ee
and two independent $R\nabla T$-type terms
\be
\label{Lag_terms_HRT}
\begin{tabular}{lll}
$H^{RT}_{3} = R^{\b\g} \divT1_{\b\g}$ ,
& $H^{RT}_{5} = R\, \trdivT1$ . 
\end{tabular}
\ee
In the kinetic nonmetricity sector, we have 16 $(\nabla Q)^2$ terms
\be
\begin{tabular}{lll}
\label{Lag_terms_HQQ}
$H^{QQ}_{1} = \nabla^\a Q^{\b\g\d}\, 
\nabla_\a Q_{\b\g\d}$ ,
& $H^{QQ}_{2} = \nabla^\a Q^{\b\g\d}\, 
\nabla_\a Q_{\g\b\d}$ ,
\\
$H^{QQ}_{3} = \nabla^\a \trQ12^\b\, \nabla_\a\trQ12_\b$ ,
& $H^{QQ}_{4} = \nabla^\a \trQ23^\b\, \nabla_\a\trQ23_\b$ ,
\\
$H^{QQ}_{5} = \nabla^\a \trQ12^\b\, \nabla_\a\trQ23_\b$ ,
\\
$H^{QQ}_{6} = \divQ1^{\a\b}\, \divQ1_{\a\b}$ ,
& $H^{QQ}_{7} = \divQ2^{\a\b}\, \divQ2_{\a\b}$ ,
\\
$H^{QQ}_{8} = \divQ2^{\a\b}\, \divQ2_{\b\a}$ ,
& $H^{QQ}_{9} = \divQ1^{\a\b}\, \divQ2_{\a\b}$ ,
\\
$H^{QQ}_{10} = %\divdivQ12^\a\, \trQ12_\a$
\divQ2^{\a\b}\nabla_\a\trQ12_\b$ ,
& $H^{QQ}_{11} = %\divdivQ12^\a\, \trQ23_\a$
\divQ2^{\a\b}\nabla_\a\trQ23_\b$ ,
\\
$H^{QQ}_{12} = %\divdivQ23^\a\, \trQ12_\a$
\divQ2^{\a\b}\nabla_\b\trQ12_\a$ ,
& $H^{QQ}_{13} = %\divdivQ23^\a\, \trQ23_\a$
\divQ2^{\a\b}\nabla_\b\trQ23_\a$ ,
\\
$H^{QQ}_{14} = (\trdivQ1)^2$ ,
& $H^{QQ}_{15} = (\trdivQ2)^2$ ,
\\
$H^{QQ}_{16} = \trdivQ1\, \trdivQ2$ ,
\end{tabular}
\ee
and four independent $R\nabla Q$ terms
\be
\begin{tabular}{lll}
\label{Lag_terms_HRQ}
$H^{RQ}_4=R^{\a\b}\, \divQ1_{\a\b}$ ,
& $H^{RQ}_5=R^{\a\b}\, \divQ2_{\a\b}$ ,
\\
$H^{RQ}_6=R\, \trdivQ1$ ,
& $H^{RQ}_7=R\, \trdivQ2$ \, . 
\end{tabular}
\ee
In the case when both torsion and nonmetricity are present, we also have 13 kinetic mixing $\nabla T\nabla Q$ terms:
\be
\begin{tabular}{lll}
\label{Lag_terms_HTQ}
$H^{TQ}_{1} = \nabla^\a T^{\b\g\d}\, 
\nabla_\a Q_{\b\g\d}$ , 
& $H^{TQ}_{2} = \nabla^\a \trT12^\b\, \nabla_\a\trQ12_\b$ , 
\\
$H^{TQ}_{3} = \nabla^\a \trT12^\b\, \nabla_\a\trQ23_\b$ , 
\\
$H^{TQ}_{4} = \divT1^{\a\b}\, \divQ1_{\a\b}$ , 
& $H^{TQ}_{5} = \divT2^{\a\b}\, \divQ2_{\a\b}$ , 
\\
$H^{TQ}_{6} = \divT1^{\a\b}\, \divQ2_{\a\b}$ , 
& $H^{TQ}_{7} = \divT1^{\a\b}\, \divQ2_{\b\a}$ , 
\\
$H^{TQ}_{8} = %\divdivT12^\a\, \trQ12_\a$
\divT2^{\a\b} \nabla_\a\trQ12_\b$ , 
& $H^{TQ}_{9} = %\divdivT12^\a\, \trQ23_\a$
\divT2^{\a\b} \nabla_\a\trQ23_\b$ , 
\\
$H^{TQ}_{10} = %\trT12_\a\, \divdivQ12^\a$
\divQ2^{\a\b} \nabla_\a\trT12_\b$ , 
& $H^{TQ}_{11} = %\trT12_\a\, \divdivQ23^\a$
\divQ2^{\a\b} \nabla_\b\trT12_\a$ , 
\\
$H^{TQ}_{12} = %D_\b \trT12^\b\, D_\a\trQ12^\a$
\trdivT1\, \trdivQ1$ , 
& $H^{TQ}_{13} = %D_\b \trT12^\b\, D_\a\trQ23^\a$
\trdivT1\, \trdivQ2$ \, . 
\end{tabular}
\ee
The numbering agrees with \eqref{LFT}, \eqref{LFQ}, \eqref{LTT}, \eqref{LQQ}, \eqref{LTQ}, while the covariant derivative associated with the independent connection is replaced with the Levi--Civita one.
In addition, we have three invariants quadratic in curvature:
\be 
\label{Lag_terms_HRR}
H^{RR}_1=R_{\mu\nu\rho\sigma}R^{\mu\nu\rho\sigma} \, , \quad
H^{RR}_2=R_{\mu\nu}R^{\mu\nu} \, , \quad
H^{RR}_3=R^2 \, .
\ee
One can use the distortion tensor instead of torsion and nonmetricity to construct equivalent bases.
It is natural to also use the Levi--Civita connection in this case.
This will give 6 independent terms of the type  $R\nabla \dis$ (denoted $H^{R\dis}_{i}$) and 38 of the type  $(\nabla \dis)^2$ (denoted $H^{\dis\dis}_{i}$).
In certain cases, they allows an even clearer representation of the results.
We list those invariants in the appendix \ref{sec:app:bases} as well for completeness.

All of them comprise the following dimension four Lagrangian contribution:
\begingroup
\allowdisplaybreaks
\begin{subequations}
\begin{align}
\label{Lag_4_Einsten}
\cL^{(\text{dim}\, 4)}_E 
=
\frac12\Big[
\sum_{i=1}^{3} b^{RR}_i H^{RR}_i
&+ \sum_{i=1}^{6} b^{R\dis}_i H^{R\dis}_i
+ \sum_{i=1}^{38} b^{\dis\dis}_i H^{\dis\dis}_i
+ \dots
%+ \textit{ints}
\Big] \, 
\\
= \label{Lag_4_EinstenRTQ}
\frac12\Big[
\sum_{i=1}^{3} b^{RR}_i H^{RR}_i
&+\sum_{i=3,5} b^{RT}_i H^{RT}_i
+\sum_{i=4,5,6,7} b^{RQ}_i H^{RQ}_i
\\ &
+\sum_{i=1}^9 b^{TT}_i H^{TT}_i
+\sum_{i=1}^{13} b^{TQ}_i H^{TQ}_i
+\sum_{i=1}^{16} b^{QQ}_i H^{QQ}_i
+ \dots
%+ \textit{ints}
\Big] \, . \nonumber
\end{align}
\end{subequations}
\endgroup
The number of independent invariants in Cartan and Einstein forms is, obviously, the same.
The relations between the two formulations have been also worked out in \cite{Baldazzi:2021kaf}, sections A.1 and A.6.

In a Riemannian manifold, there exists a combination that comprises a topological invariant after integration over the whole space-time volume, called Euler invariant or Gau\ss-Bonnet term $H^{RR}_1 - 4 H^{RR}_2 + H^{RR}_3$.
Its Cartan space analog is 
\begin{subequations}
\begin{align}
\label{EulerInvariantCartan}
&\int d^4 x \sqrt{g}\, E_{GB} 
\nonumber\\ 
= &\int d^4 x \sqrt{g}\, \left[ L^{FF}_3 - 4 L^{FF}_7 + L^{FF}_{16} \right]
\\
= &\int d^4 x \sqrt{g}\, \left[ H^{RR}_{1} - 4 H^{RR}_{2} + H^{RR}_{3} + 4 H^{R\dis}_{1} + 8 H^{R\dis}_{4} - 8 H^{R\dis}_{8} + 2 H^{R\dis}_{11} - 2 H^{R\dis}_{12} 
\right. \nonumber\\ & \left.
- 4 H^{\dis\dis}_{6} - 4 H^{\dis\dis}_{14} + H^{\dis\dis}_{15} + H^{\dis\dis}_{17} - 2 H^{\dis\dis}_{23} + 8 H^{\dis\dis}_{24} + H^{\dis\dis}_{34} + H^{\dis\dis}_{35} - 2 H^{\dis\dis}_{38} 
\right] \, .
\end{align}
\end{subequations}

The action we consider in this paper is \eqref{lag_XYZ_ours}.
We can now present it in our contracted notations in Cartan and Einstein forms as
\begingroup
\allowdisplaybreaks
\begin{subequations}
\begin{align}
\label{Lag_XYZ_Cartan}
S 
= & - \int d^4 x \sqrt{g} \left[ - a_0 F + L^{FF}_1 + M^{TT}_1 + M^{QQ}_1 \right]
\\ 
= & - \int d^4 x \sqrt{g} \left[ c_1 \left\{
H^{RR}_{1} + 4 H^{R\dis}_{1} + 2 H^{\dis\dis}_{1} - 2 H^{\dis\dis}_{12} \right\} 
\right. \nonumber\\ & \left.
+ 2 (a_1 + a_4) M^{\dis\dis}_1
+ 2 a_4 M^{\dis\dis}_2 - 2 a_1 M^{\dis\dis}_3 + a_0 \left\{ M^{\dis\dis}_5 - M^{\dis\dis}_9 - R \right\} 
\right] \label{lag_XYZ_Einstein}
\\ \label{Lag_XYZ_EinsteinRTQ}
= & - \int d^4 x \sqrt{g} \left[ 
c_1 \left\{ H^{RR}_{1} + 8 H^{RT}_{3} - 4 H^{RQ}_{1} + \frac{3}{2} H^{TT}_{1} + H^{TT}_{2} - H^{TT}_{4} + H^{TT}_{5} - \frac{1}{2} H^{TT}_{6} \right. \right. 
\nonumber\\ & \left.
- 2 H^{TT}_{7} 
- 4 H^{TQ}_{1} + 2 H^{TQ}_{5} + 2 H^{TQ}_{6} - 2 H^{TQ}_{7} + \frac{3}{2} H^{QQ}_{1} - H^{QQ}_{2} - \frac{1}{2} H^{QQ}_{6} - H^{QQ}_{7} + H^{QQ}_{8} \right\} 
\nonumber \\ & 
+ (a_1 - \frac{a_0}{4}) m^{TT}_{1}
+ (a_4 - \frac{a_0}{4}) m^{QQ}_{1}
+ a_0 \left\{ - \frac{1}{2} m^{TT}_{2} + m^{TT}_{3} 
\right. \nonumber \\ & \left. \left.
+ m^{TQ}_{1} + m^{TQ}_{2} - m^{TQ}_{3} 
 + \frac{1}{2} m^{QQ}_{2} + \frac{1}{4} m^{QQ}_{3} - \frac{1}{2} m^{QQ}_{5} - R
 \right\} \right] \, . 
\end{align}
\end{subequations}
\endgroup
We will use the first form \eqref{Lag_XYZ_Cartan} to perform perturbative calculation and the other two for some considerations later.

%%%%%%%%%%%
\section{\label{sec:loop_calculation}Setup for one loop calculation}
%%%%%%%%%%%

The calculations have been performed in the Euclidean regime,
where the action differs from \eqref{lag_XYZ_ours}, \eqref{Lag_XYZ_Cartan} by an overall sign.

%%%%%%%%%%%
\subsection{\label{subsec:expansion}Perturbative Expansion}
%%%%%%%%%%%

Let us consider now perturbations in generic frames:
\begin{subequations}
\begin{align}
\theta^{a}{}_{\m}{} &= \bar{\theta}^{a}{}_{\m}{} + {X}^{a}{}_{\m}{} \, , 
\\ 
g_{ab} &= \bar{g}_{ab} + {Y}_{ab} \, , 
\\
A_{\m}{}^{a}{}_{b}{} &= \bar{A}_{\m}{}^{a}{}_{b}{} + { Z}_{\m}{}^{a}{}_{b}{} \, . 
\end{align}
\end{subequations}
Here and below the bars denote generic background values.
Then from \eqref{pullbackmetric}, \eqref{pullbackconnection} we have for the pullback metric and connection on the tangent bundle:
\begin{subequations}
\begin{align}
\label{ginXYZ}
{\d g}_{\m \n } &= X_{\m \n } + X_{\n \m } + Y_{\m \n } + X_{a\n } X^{a}{}_{\m } + X^{a}{}_{\n } Y_{\m a} + X^{a}{}_{\m } Y_{\n a} + \dots \, , 
\\ \label{AinXYZ}
\d A_{\m }{}^{\a }{}_{\b } &= Z_{\m }{}^{\a }{}_{\b } + \bar{\nabla }_{\m } X^{\a }{}_{\b } - X^{\a }{}_{\s} Z_{\m }{}^{\s}{}_{\b } + X^{\s}{}_{\b } Z_{\m }{}^{\a }{}_{\s} - X^{\a }{}_{\s} \bar{\nabla}_{\m }X^{\s}{}_{\b } + \dots \, ,
\end{align}
\end{subequations}
where the bars over the covariant derivatives indicate that they are computed with the background metric.
For the perturbations of the inverse metric and metric determinant, we have
\begin{subequations}
\begin{align}
g^{\m \n } &= \bar{g}^{\m \n } - X^{\m \n } - X^{\n \m } - Y^{\m \n } \nonumber \\ 
&+ X_{a}{}^{\n } X^{\m a} + X^{a\m } X^{\n }{}_{a} + X^{\m a} X^{\n }{}_{a} + X^{\n a} Y_{a}{}^{\m } + X^{\m a} Y_{a}{}^{\n } + Y_{a}{}^{\n } Y^{\m a} + \dots \, ,
\\
\sqrt{- g} &= \sqrt{- \bar{g}} \left( 1 + X^{\a}{}_{\a} + \frac{1}{2} Y^{\a}{}_{\a} \right. \nonumber \\
&- \left. \frac{1}{2} X_{\a\b} X^{\b\a} + \frac{1}{2} X^{\a}{}_{\a} X^{\b}{}_{\b} - \frac{1}{4} Y_{\a\b} Y^{\b\a} + \frac{1}{2} X^{\a}{}_{\a} Y^{\b}{}_{\b} + \frac{1}{8} Y^{\a}{}_{\a} Y^{\b}{}_{\b} \right) + \dots \, .
\end{align}
\end{subequations}
In the following, we will need the commutator of the covariant derivatives. 
If the covariant derivative is the general one, we have
\bear
\label{commutnoncompat}
\left[ \bar{D}_{\t}, \bar{D}_{\s}\right] X^{\a }{}_{\b } &= \d_{\b }{}^{\d } \bar{F}_{\t \s }{}^{\a }{}_{\g } X^{\g }{}_{\d } - \d_{\g }{}^{\a } \bar{F}_{\t \s }{}^{\d }{}_{\b } X^{\g }{}_{\d } - \bar{T}_{\t}{}^{\l}{}_{\s} \bar{D}_{\l} X^{\a }{}_{\b }, 
\\
\left[ \bar{D}_{\t}, \bar{D}_{\s} \right] Y_{\a \b } &= - \d_{\b }{}^{\d } \bar{F}_{\t \s }{}^{\g }{}_{\a } Y_{\g \d } - \d_{\a }{}^{\g } \bar{F}_{\t \s}{}^{\d }{}_{\b } Y_{\g \d } - \bar{T}_{\t}{}^{\l}{}_{\s} \bar{D}_{\l} Y_{\a \b }, 
\\
\left[ \bar{D}_{\t}, \bar{D}_{\s} \right] Z_{\a }{}^{\b }{}_{\g } &= - \d_{\g }{}^{\d } \d_{\eta }{}^{\b } \bar{F}_{\t \s }{}^{\zeta }{}_{\a } Z_{\z }{}^{\eta }{}_{\d } + \d_{\a }{}^{\zeta } \d_{\g }{}^{\d } \bar{F}_{\t \s }{}^{\b }{}_{\eta } Z_{\zeta }{}^{\eta }{}_{\d } 
\\ 
&- \d_{\a }{}^{\z } \d_{\eta }{}^{\b } \bar{F}_{\t \s }{}^{\d }{}_{\g } Z_{\z }{}^{\eta }{}_{\d } - \bar{T}_{\t}{}^{\l}{}_{\s} \bar{D}_{\l} Z_{\a }{}^{\b }{}_{\g }.
\eear
Notice the appearance of terms linear in torsion and derivatives.
For the LC covariant derivatives, we get the corresponding formulas by replacing the general curvature tensor $F$ with the Riemann tensor $R$ and putting the torsion to zero.
In this case, the expression for the commutator can be factorized: 
\bear
\label{OmegaXYZdef}
\left[ \bar{\na}_{\t}, \bar{\na}_{\s} \right] X^{\a }{}_{\b } &= \Omega^{X}{}_{\t \s }{}^{\a }{}_{\b \g }{}^{\d } X^{\g }{}_{\d }, 
\\
\left[ \bar{\na}_{\t}, \bar{\na}_{\s} \right] Y_{\a \b } &= \Omega^{Y}{}_{\t \s \a \b }{}^{\g \d } Y_{\g \d }, 
\\
\left[ \bar{\na}_{\t}, \bar{\na}_{\s} \right] Z_{\a }{}^{\b }{}_{\g } &= \Omega^{Z}{}_{\t \s \a }{}^{\b }{}_{\g }{}^{\zeta }{}_{\eta }{}^{\d } Z_{\zeta }{}^{\eta }{}_{\d }, 
\eear
where
\bear
\label{OmegaXYZexpression}
\Omega^{X}{}_{\t \s }{}^{\a }{}_{\b \g }{}^{\d } &= \d_{\b }^{\d } \bar{R}_{\t \s }{}^{\a }{}_{\g } - \d_{\g }^{\a } \bar{R}_{\t \s }{}^{\d }{}_{\b }, 
\\
\Omega^{Y}{}_{\t \s \a \b }{}^{\g \d } &= - \d_{\b }^{\d } \bar{R}_{\t \s }{}^{\g }{}_{\a } - \d_{\a }^{\g } \bar{R}_{\t \s}{}^{\d }{}_{\b }, 
\\
\Omega^{Z}{}_{\t \s \a }{}^{\b }{}_{\g }{}^{\zeta }{}_{\eta }{}^{\d } &= - \d_{\g }^{\d } \d_{\eta }^{\b } \bar{R}_{\t \s }{}^{\zeta }{}_{\a } + \d_{\a }^{\zeta } \d_{\g }^{\d } \bar{R}_{\t \s }{}^{\b }{}_{\eta } - \d_{\a }^{\zeta } \d_{\eta }^{\b } \bar{R}_{\t \s }{}^{\d }{}_{\g }. 
\eear
This can be expressed in condensed notations as
\be
\left[ \bar{\na}_{\t}, \bar{\na}_{\s} \right] \psi = \mathbb{\Omega}\; \psi \, ,
\ee
where $\psi = \left( X,Y,Z \right)^T$ and 
\be
\label{OmegaXYZmatrix}
\mathbb{\Omega} = 
\begin{pmatrix}
\Omega^{X}{}_{\t \s }{}^{\a_1 }{}_{\b_1 \g_1 }{}^{\d_1 } & 0 & 0 \\
0 & \Omega^{Y}{}_{\t \s \a_2 \b_2 }{}^{\g_2 \d_2 } & 0 \\
0 & 0 & \Omega^{Z}{}_{\t \s \a_3 }{}^{\b_3 }{}_{\g_3 }{}^{\zeta_3 }{}_{\eta_3 }{}^{\d_3 } 
\end{pmatrix} \, .
\ee
Using (\ref{commutnoncompat}), we obtain the following perturbative expansions for the curvature, torsion, and nonmetricity tensors:
\begingroup
\allowdisplaybreaks
\begin{subequations}
\begin{align}
F_{\m \n }{}^{\r }{}_{\l } &= \bar{F}_{\m \n }{}^{\r }{}_{\l } + \bar{F}_{\m \n }{}^{\r }{}_{\s } X^{\s }{}_{\l } - \bar{F}_{\m \n }{}^{\s }{}_{\l } X^{\r }{}_{\s } + \bar{F}_{\m \n }{}^{\t }{}_{\l } X^{\s }{}_{\t } X^{\r }{}_{\s } - \bar{F}_{\m \n }{}^{\s }{}_{\t } X^{\t }{}_{\l } X^{\r }{}_{\s } - X^{\r }{}_{\s } \bar{D }_{\m }Z_{\n }{}^{\s }{}_{\l }
\nn \\ 
& + X^{\s }{}_{\l } \bar{D }_{\m }Z_{\n }{}^{\r }{}_{\s } + \bar{D }_{\m }Z_{\n }{}^{\r }{}_{\l } + X^{\r }{}_{\s } \bar{D }_{\n }Z_{\m }{}^{\s }{}_{\l } - X^{\s }{}_{\l } \bar{D }_{\n }Z_{\m }{}^{\r }{}_{\s } - \bar{D }_{\n }Z_{\m }{}^{\r }{}_{\l } + \nn \\
 &+ Z_{\m }{}^{\r }{}_{\s } Z_{\n }{}^{\s }{}_{\l } - Z_{\m }{}^{\s }{}_{\l } Z_{\n }{}^{\r }{}_{\s } 
+ \bar{T}_{\m }{}^{\s }{}_{\n } \left(Z_{\s }{}^{\r }{}_{\l}- X^{\r }{}_{\t } Z_{\s }{}^{\t }{}_{\l } + X^{\t }{}_{\l } Z_{\s }{}^{\r }{}_{\t } \right) + \dots, 
\\
Q_{\l \m \n } & = \bar{Q}_{\l \m \n } - \bar{D }_{\l }Y_{\m \n } + \bar{Q}_{\l \s \n } X^{\s }{}_{\m } + \bar{Q}_{\l \m \s } X^{\s }{}_{\n } + \bar{Q}_{\l \s \t } X^{\s }{}_{\m } X^{\t }{}_{\n } + 
\nn \\ 
& + g_{\s \n } Z_{\l }{}^{\s }{}_{\m } + g_{\s \m } Z_{\l }{}^{\s }{}_{\n } + g_{\t \n } X^{\s }{}_{\m } Z_{\l }{}^{\t }{}_{\s } + g_{\t \m } X^{\s }{}_{\n } Z_{\l }{}^{\t }{}_{\s } \nn \\ 
& + g_{\t \s } X^{\s }{}_{\n } Z_{\l }{}^{\t }{}_{\m } + g_{\t \s } X^{\s }{}_{\m } Z_{\l }{}^{\t}{}_{\n } - 
 -X^{\s }{}_{\m } \bar{D }_{\l }Y_{\s \n } - X^{\s }{}_{\n } \bar{D }_{\l }Y_{\m \s } + \dots
\\
T_{\m }{}^{\l }{}_{\n } &= \bar{T}_{\m }{}^{\l }{}_{\n } + Z_{\m }{}^{\l }{}_{\n } - Z_{\n }{}^{\l }{}_{\m } - X^{\l }{}_{\s} Z_{\m }{}^{\s}{}_{\n } + X^{\l }{}_{\s} Z_{\n }{}^{\s}{}_{\m } +
\nn \\ 
& + \bar{D}_{\m }X^{\l }{}_{\n } - \bar{D}_{\n }X^{\l }{}_{\m } + X^{\l }{}_{\s} \bar{D}_{\n }X^{\s}{}_{\m } - X^{\l }{}_{\s} \bar{D}_{\m }X^{\s}{}_{\n } + \dots 
\end{align}
\end{subequations}
\endgroup
Performing the second variation of the action \eqref{lag_XYZ_ours} using these relations one arrives at the Hessian, which we write in coordinate bases, by converting all Latin indices to Greek ones for convenience.
The principal part is second order in derivatives:
\be \label{Hessian_nonmin}
a_1 X_{\m\n}(-\bar\Box g^{\n\sigma}+\bar\nabla^\sigma\bar\nabla^\n)X^\m{}_{\sigma} 
+ a_4 Y_{\m\n}(-\bar\Box g^{\n\sigma}) Y^{\m\n}
+c_1 Z_\m{}^{\r\sigma}(-\bar\Box g^{\m\n}+\bar\nabla^\n\bar\nabla^\m)Z_{\n\r\sigma} \, ,
\ee
and it contains a minimal part (proportional to $\bar \Box = \bar \nabla^\m \bar \nabla_\m$ and identity operator in the space of fields), and two nonminimal parts.
The latter would greatly complicate the calculation, but as we will see below, they can be avoided by choosing a suitable gauge.
This is only possible because we started in the formalism of generic frames.
The Hessian vanishes on fields that are local Lorentz or diffeomorphism transformations.
Starting in coordinate bases, one would have to fix only 
the diffeomorphism invariance
(4 parameters) and one can fix this gauge in such a way as to remove the nonminimal term in the $X$-$X$ sector.
By choosing a suitable $GL(4)$ gauge fixing (16 parameters) we can remove also the nonminimal contribution in the $Z$-$Z$ sector.
Working with the full gauge group $\mathbb{R}^{1, 3} \rtimes GL(4)$ allows us to remove both of them.
Such simplification comes with the cost of a more complicated gauge fixing procedure, which the next subsection is devoted to.
It largely repeats the considerations of \cite{Melichev:2023lwj} (with a difference of a larger gauge group), we keep it here for completeness.
For similar discussions in MAGs and Yang--Mills theory we refer to \cite{Jackiw:1978ar,
Daum:2009dn, Daum:2010qt,
Benedetti:2011nd,
Daum:2010zz,
Daum:2013fu}.

%%%%%%%%%%%
\subsection{\label{subsec:gauge_algebra}Gauge algebra}
%%%%%%%%%%%

Consider infinitesimal transformations \eqref{transform_GL4_and_diffeos} with
\begin{subequations}
 \begin{align}
\label{diffeos_infinitesimal}
x'^\m &= x^\m - \x^\m \, .
\\ \label{GL4_infinitesimal}
\Lambda^{a}{}_b &= \mathbb{1}^{a}{}_b + \omega^{a}{}_b \, .
 \end{align}
\end{subequations}
We have
\bear \label{GL4_acts}
\d^{\text{L}}_\omega\theta^a{}_\m &= -\omega^a{}_b\theta^b{}_\m \, ,
\\
\d^{\text{L}}_\omega g_{ab} &= \omega_{ab} + \omega_{ba} \, ,
\\
\d^{\text{L}}_\omega A_\m{}^a{}_b &= D_\m\omega^a{}_b \, ,
\eear
for the infinitesimal $GL(4)$ transformations and
\bear \label{diffeos_act}
\d^{\text{D}}_\x\theta^a{}_\m & = \cL_\x \theta^a{}_\m
= \x^\r\partial_\r\theta^a{}_\m + \left( \partial_\m \x^\r \right) \theta^a{}_\r \, , 
\\
\d^{\text{D}}_\x g_{ab} & = \cL_\x g_{ab}= \x^\r \partial_\r g_{ab} \, ,
\\
\d^{\text{D}}_\x A_\m{}^a{}_b & = \cL_\x A_\m{}^a{}_b = \x^\r \partial_\r A_\m{}^a{}_b + \left( \partial_\m \x^\r \right) A_\r{}^a{}_b \, .
\eear
for the infinitesimal diffeomorphisms, where $\cL_\x$ is the Lie derivative.
Note that the Latin indices are inert under this definition of diffeomorphism.
The algebra of these transformations is
\bear
\left[\d^{\text{L}}_{\omega_1},\d^{\text{L}}_{\omega_2}\right]&=\d^{\text{L}}_{[\omega_1,\omega_2]} \, ,
\\
\left[\d^{\text{D}}_{\x_1},\d^{\text{D}}_{\x_2}\right]&=-\d^{\text{D}}_{[\x_1,\x_2]} \, ,
\\
\left[\d^{\text{D}}_{\x},\d^{\text{L}}_{\omega}\right]&=\d^{\text{L}}_{\cL_\x\omega} \, .
\eear
This shows that the local $GL(4)$ transformations are a normal subgroup of the full gauge group, and the diffeomorphisms are the quotient
of the full group by this subgroup.

Now we observe that whereas the general soldering fluctuation $X^a{}_\m$ transforms properly
under the $GL(4)$ transformations, the gauge fluctuation $\d_\x\theta^a{}_\m$ 
does not.
This would become a serious obstacle in the following, 
because $\d_\x\theta^a{}_\m$ is used in the
construction of the ghost operator, and this definition would lead to a
non-covariant ghost operator.
\be
\tilde \d^{\text{D}}_\x=\d^{\text{D}}_\x-\d^{\text{L}}_{\x\cdot A} \, . 
\label{modified}
\ee
which additionally shifts the diffeomorphisms by a parameter
$\epsilon^a{}_b = -\x^\m A_\m{}^a{}_b\equiv -(\x\cdot A)^a{}_b$, but the normal subgroup remains untouched.
The action of these modified diffeomorphisms on the fields is
\bear \label{modified_diffeos_act}
\tilde\d^{\text{D}}_\x\theta^a{}_\m&=
\theta^a{}_\r\nabla_\m \x^\r
+\x^\r \dis_\r{}^a{}_\m
 \, ,
\\
\tilde\d^{\text{D}}_\x g_{ab} &= - \x^\r Q_{\r ab} \, ,
\\
\tilde\d^{\text{D}}_\x A_\m{}^a{}_b&=\x^\r F_{\r\m}{}^a{}_b \, ,
\eear
where we used (\ref{totcovd_of_soldering}).
Defining $F(\x_1,\x_2)^a{}_b=\x_1^\m \x_2^\n F_{\m\n}{}^a{}_b$ we obtain for the gauge algebra
\bear
\left[\tilde\d^{\text{D}}_{\x_1},\tilde\d^{\text{D}}_{\x_2}\right]&=
-\tilde\d^{\text{D}}_{[\x_1,\x_2]}-\d^{\text{L}}_{F(\x_1,\x_2)} \, ,
\\
\left[\tilde\d^{\text{D}}_{\x},\d^{\text{L}}_{\omega}\right]&=0 \, .
\label{modalg}
\eear

In background field calculations, one has to split the action of the gauge transformations of the full field into the transformations of the background and of the perturbations.
Following the standard nomenclature, we define the \textbf{``quantum''} transformations $\d^{\text{Q}}$ the backgrounds are invariant,
and the whole transformation of the field is attributed to the fluctuation.
For the general linear transformations, this means
\bear
\d^{\text{QL}}_\omega \bar\theta^a{}_\m & = 0 \, ,
\\
\d^{\text{QL}}_\omega \bar g_{ab} & = 0 \, ,
\\
\d^{\text{QL}}_\omega \bar A_\m{}^a{}_b & = 0 \, ,
\\
\d^{\text{QL}}_\omega X^a{}_\m & = -\omega^a{}_b\,\theta^b{}_\m \, ,
\\
\d^{\text{QL}}_\omega Y_{ab} & = \omega_{ab} + \omega_{ba} \ ,
\\
\d^{\text{QL}}_\omega Z_\m{}^a{}_b & = \bar D_\m\omega^a{}_b \, ;
\eear
and for the diffeomorphisms we define
\bear
\tilde\d^{\text{QD}}_\x\bar\theta^a{}_\m & = 0 \, ,
\\
\tilde\d^{\text{QD}}_\x\bar g_{ab} & = 0 \, ,
\\
\tilde\d^{\text{QD}}_\x \bar A_\m{}^a{}_b & = 0 \, ,
\\
\tilde\d^{\text{QD}}_\x X^a{}_\m & = 
\theta^a{}_\r\bar\nabla_\m v^\r+v^\r \bar\dis_\r{}^a{}_\m \, ,
\\
\tilde\d^{\text{QD}}_\x Y^a{}_\m & = 
- v^\r \bar Q_{\r ab} \, ,
\\
\tilde\d^{\text{QD}}_\x Z_\m{}^a{}_b & = v^\r \bar F_{\r\m}{}^a{}_b \, .
\eear
The \textbf{``background''} transformations $\d^{\text{B}}$ are defined in such a way that
the backgrounds transform
as the original field (in particular, $\bar A$ transforms as a connection).
In detail, the background $GL(4)$ transformations are
\bear
\d^{\text{BL}}_\omega\bar\theta^a{}_\m & = 
-\omega^a{}_b \,\bar\theta^b{}_\m
 \, ,
\\
\d^{\text{BL}}_\omega\bar g_{ab} & = 
\omega_{ab} + \omega_{ba}
 \, ,
\\
\d^{\text{BL}}_\omega \bar A_\m{}^a{}_b & = \bar D_\m\omega^a{}_b \, ,
\\
\d^{\text{BL}}_\omega X^a{}_\m & = 
-\omega^a{}_c \,X^c{}_\r
 \, ,
\\
\d^{\text{BL}}_\omega Y_{ab} & = 
0
 \, ,
\\
\d^{\text{BL}}_\omega Z_\m{}^a{}_b & = 
-\omega^a{}_c\, Z_\m{}^c{}_b
+Z_\m{}^a{}_c \,\omega^c{}_b
 \, ;
\eear
and the background diffeomorphisms are given by the Lie derivative on all fields.
The background diffeomorphisms can be covariantized as above, in particular
\bear
\bar\d^{\text{BD}}_\x\bar\theta^a{}_\m & = \bar\theta^a{}_\r\bar\nabla_\m v^\r
+v^\r \bar \dis_\r{}^a{}_\m \, ,
\\
\bar\d^{\text{BD}}_\x \bar A_\m{}^a{}_b & = v^\r \bar F_{\r\m}{}^a{}_b \, .
\eear

%%%%%%%%%%%
\subsection{\label{subsec:gauge_fixed_Hessian}Gauge fixed Hessian}
%%%%%%%%%%%

We fix the gauge by choosing the Lorentz-like gauge conditions
\begin{subequations}
\begin{align}
\chi_{\text{D}}^\m & = \bar\nabla^\n X^\m{}_\n \, ,
\\
\chi_{\text{L}}{}^a{}_b & = \bar\nabla^\n Z_\n{}^a{}_b \, .
\end{align}
\end{subequations}
In the latter expression, it is understood that the covariant derivative is defined in terms of the background LC connection for both types of indices.
Then, the gauge fixing action is
\be
S_{gf}= \polov \int d^4x\sqrt{\bar g}\left[
\frac{a_1}{\a_D} \bar g_{\r\sigma}\chi_{\text{D}}^\r\chi_{\text{D}}^\sigma
+\frac{c_1}{\a_L}\bar g_{ac}\bar g^{bd}
\chi_{\text{L}}{}^a{}_b \chi_{\text{L}}{}^c{}_d\right] \, . 
\ee
This breaks invariance under the ``quantum'' transformations
while preserving invariance under the ``background'' transformations.
The total background covariant derivative of the background soldering is zero, making it simple to transform all the Latin indices into Greek ones.
We set the parameters $\a_D=\a_L=1$ (Feynman gauge), integrating by parts and commuting derivatives we get
\bea
S_{gf}&=& \polov \int d^4x\sqrt{\bar g}\Big[
-a_1 X_{\m\n}\bar\nabla^\sigma\bar\nabla^\n X^\m{}_\sigma
-c_1 Z_\m{}^{\r\sigma}\bar\nabla^\n\bar\nabla^\m Z_{\n\r\sigma}
\label{gaugefixing}
\\
&&
+a_1 X_{\m\n}\bar R^{\n\sigma}X^\m{}_\sigma
-a_1 X_{\m\n}\bar R^{\m\r\n\sigma}X_{\r\sigma}
+c_1 Z_{\m\a\b}\bar R^{\m\sigma}Z_\sigma{}^{\a\b}
-2c_1 Z_{\m\a\b}\bar R^{\m\r\a\sigma}Z_{\r\sigma}{}^{\b}
%+Z_{\m\a\b}\bar R^{\m\r\b\sigma}Z_{\r\a\sigma}
\Big] \, . 
\nonumber
\eea
The first line exactly cancels the unwanted nonminimal terms in (\ref{Hessian_nonmin}).
At this moment it is convenient to rescale the kinetic variables 
\be
X^\a{}_\m \to \frac{1}{\sqrt{2 a_1}}X^\a{}_\m \, ,\qquad
Y_{\a\b} \to \frac{1}{\sqrt{a_4}} Y_{\a\b} \, ,\qquad
Z_\m{}^\a{}_\b\to \frac{1}{\sqrt{2 c_1}}Z_\m{}^\a{}_\b \, .
\label{rescaling}
\ee
Performing some integrations by parts, one can write the gauge-fixed Hessian in the form
\begin{equation}
H = \frac12(\Psi, \cO \Psi) \, .
\end{equation}
where $\Psi = (X, Y, Z)^T$ and
\be
\label{KinopGeneral}
\cO = -\bar\Box \mathbb{1} + \mathbb{V}^\sigma \bar\nabla_\sigma+ \mathbb{W} \ ,
\ee
where the $\mathbb{V}$ and $\mathbb{W}$ are matrices in the space of the fields
\be
 \mathbb{V}^\m=
\begin{pmatrix}
V_{XX}^\m & V_{XY}^\m & V_{XZ}^\m
\\
V_{YX}^\m & V_{YY}^\m & V_{YZ}^\m
\\
V_{ZX}^\m & V_{ZY}^\m& V_{ZZ}^\m
\end{pmatrix} \, ,
\qquad\qquad
\mathbb{W}=
\begin{pmatrix}
W_{XX} & W_{XY} & W_{XZ}
\\
W_{YX} & W_{YY} & W_{YZ}
\\
W_{ZX} & W_{ZY} & W_{ZZ} 
\end{pmatrix} \, ,
\ee
with the appropriate free indices.
%Note that this contains all the terms coming from the second variation of the action, plus the terms in the second line of (\ref{gaugefixing}).
With the rescaling (\ref{rescaling}) the quantum fields $X$, $Y$ and $Z$ have canonical dimension 1, $\mathbb{V}$ has dimension 1 and $\mathbb{W}$ has dimension 2.
At the high energy scale $E$ matrices $\mathbb{V}$ and $\mathbb{W}$ can be treated as small perturbations
compared to the leading $\bar\nabla^2\approx E^2$ term under the following assumptions about the couplings
\be
\frac{a_1}{c_1}\ll E^2 \, ,\quad \frac{a_4}{c_1}\ll E^2 \, ,
\ee
and the background
\be
\bar T\ll E \, ,\quad 
\bar\nabla \bar T\ll E^2 \, ,\quad
\bar Q\ll E \, ,\quad
\bar\nabla \bar Q\ll E^2 \, , \quad 
\bar F\ll\sqrt{\frac{a_1}{c_1}}E\ll E^2 \, .
\ee
These conditions are necessary for the applicability of the
early-time asymptotic heat kernel expansion.

%%%%%%%%%%%%%%%%%%%%%%%%%%%%%%%%%%

Let us mention that it is owing to the particular choice of the action \eqref{lag_XYZ_ours} and the gauge-fixing conditions and parameters that we managed to arrive at the second-order kinetic operator with a minimal principal part \eqref{KinopGeneral}.
Considerations of \cite{Melichev:2023lpn}, section 8.1 show that it does not seem to be possible to perform a similar trick in general.
Indeed, let us look at the following Lagrangian:
\be
\cL = \frac{1}{2} \left[ c_1 F_{\m\n\r\l} F^{\m\n\r\l} + b_1 \nabla_\m T_{\n\r\l} \nabla^\m T^{\n\r\l} \right] \, .
\ee
For simplicity, let us assume that the nonmetricity vanishes.
Then, the principal part of the Hessian (written in a self-adjoint form) is
\bear \label{kinop_DTsquared}
&b_1 X^{\m\n} \bar\Box^2 X_{\m\n} - (b_1+c_1) Z^{\m\n\r} \bar\Box Z_{\m\n\r} + b_1 Z^{\m\n\r} \bar\Box Z_{\n\m\r} - b_1 X^{\m\n} \bar\Box \bar\nabla_\r Z_{\n\m}{}^\r 
\\
&+ b_1 X^{\m\n} \bar\Box \bar\nabla_\r Z^\r{}_{\m\n} - b_1 X^{\m\n} \bar\Box \bar\nabla_\n \bar\nabla_\r X_\m{}^\r + b_1 Z^{\m\n\r} \bar\Box \bar\nabla_\r X_{\n\m} - b_1 Z^{\m\n\r} \bar\Box \bar\nabla_\m X_{\n\r} 
\\
&+ c_1 Z^{\m\n\r} \bar\nabla_\m \bar\nabla_\l Z^\l{}_{\n\r} \, .
\eear
Consider a family of gauge-fixing conditions 
\bear \label{gauge-fix_cond-general schematic}
\chi_{\text{D}} &= \bar\nabla X + Z \, ,
\\
\chi_{\text{L}} &= \bar\nabla Z + \bar\nabla \bar\nabla X + X \, .
\eear
We write it here in a schematic way, suppressing indices that can be contracted in various ways.
These conditions can be enforced by the action contribution of the form
%\footnote{In principle, one could consider even more general forms containing more derivatives but this is irrelevant in our case of the fourth-order kinetic operator \eqref{kinop_DTsquared}.}
\bear \label{gauge-fix_term-general schematic}
S_{gf} =& \frac{1}{2\a} \int d^4x\sqrt{\bar g}\left[
\bar g_{\r\sigma}\chi_{\text{D}}^ \r\chi_{\text{D}}^\sigma
+ \z_1 \bar g_{\r\sigma} \chi_{\text{D}}^\r \bar \Box \chi_{\text{D}}^\r
+ \z_2 \chi_{\text{D}}^\r \bar\nabla_\r \bar\nabla_\s \chi_{\text{D}}^\s
\right.\\
&\left. +\x_1 \bar g_{ac}\bar g^{bd}
\chi_{\text{L}}{}^a{}_b \chi_{\text{L}}{}^c{}_d
+\x_2 \chi_{\text{L}}{}^a{}_b \chi_{\text{L}}{}^b{}_a +\x_3 \chi_{\text{L}}{}^a{}_a \chi_{\text{L}}{}^b{}_b + O ( \bar R )
\right] \, . 
\eear
Looking at various possible expressions of the form \eqref{gauge-fix_cond-general schematic}
% as well as various choices for the coefficients $\z$ $\x$ in \eqref{gauge-fix_term-general schematic} 
one can convince oneself of the impossibility of rendering the operator \eqref{kinop_DTsquared} minimal.
In the following, we stick to the Lagrangian \eqref{lag_XYZ_ours} until a brief discussion of the general picture in section \ref{sec:Discussion}.

%%%%%%%%%%%%%%%%%%%%%%%%%%%
\subsection{Ghost action}
%%%%%%%%%%%%%%%%%%%%%%%%%%%

According to the Faddeev--Popov procedure, we need to define the ghost action. 
The ghost operators are the infinitesimal ``quantum'' transformations applied to the gauge fixing conditions, with the transformation parameters $\omega^a{}_b$ and $\x^\m$ replaced by the ghost fields $\Sigma^a{}_b$ and $\pi^\m$.
We define 
\be
\d^{\text{QL}}_\Sigma\chi_{\text{L}}=\d_{LL}\Sigma\ ,\quad
\tilde\d^{\text{QD}}_{\pi}\chi_{\text{L}}=\d_{LD} \pi\ ,\quad
\d^{\text{QL}}_\Sigma\chi_{\text{D}}=\d_{DL}\Sigma\ ,\quad
\tilde\d^{\text{QD}}_{\pi}\chi_{\text{D}}=\d_{DD} \pi\ ,
\label{ghostops}
\ee
and then the ghost action is then given by
\bear
S_{gh}&=\int d^4x\sqrt{\bar g}
\left[
\bar\Sigma(\d^{\text{QL}}_\Sigma\chi_{\text{L}}+\tilde\d^{\text{QD}}_{\pi} \chi_{\text{L}})
+\bar {\pi}(\d^{\text{QL}}_\Sigma\chi_{\text{D}}+\tilde\d^{\text{QD}}_{\pi} \chi_{\text{D}})
\right]
\\
&=
\int d^4x\sqrt{\bar g}
\begin{pmatrix}
\bar\Sigma& \bar \pi
\end{pmatrix}
\begin{pmatrix}
\d_{LL} & \d_{LD}\\
\d_{DL} & \d_{DD}
\end{pmatrix}
\begin{pmatrix}
\Sigma\\ \pi
\end{pmatrix} \, ,
\eear
where $\bar\Sigma_a{}^b$ and $\bar \pi_\m$ are the antighost fields.
We suppressed the indices for clarity.
%All indices here have been suppressed for clarity.
Explicitly evaluating the infinitesimal transformations in (\ref{ghostops}), we would obtain operators of the form $\bar\nabla\nabla$, \ie containing both the background and the full connection.
However, we are ultimately interested only in the effective action at zero fluctuation fields, so we can replace the full fields with their background.
The ghost operators are then constructed entirely with
background fields.
Due to the fact that the total covariant derivative of the background soldering with the background LC connection is zero, we can write all formulas using only coordinate (Greek) indices, without producing new terms.
The ghost operators are then:
\bear \label{ghost_op}
(\d_{LL}\Sigma)^\a{}_\b &= \bar\nabla^2\Sigma^\a{}_\b
+\bar\nabla^\n(\bar \dis_\n{}^\a{}_\g\Sigma^\g{}_\b
-\Sigma^\a{}_\g\bar \dis_\n{}^\g{}_\b)
\\
(\d_{LD} \pi)^\a{}_\b &=
\bar\nabla^\n(\bar F_{\r\n}{}^\a{}_\b \pi^\r)
\\
(\d_{DL}\Sigma)^\a &= -\bar\nabla^\n\Sigma^\a{}_\n
\\
(\d_{DD} \pi)^\a &= \bar\nabla^2 \pi^\a
+\bar\nabla^\n(\bar \dis_\r{}^\a{}_\n \pi^\r) \, . 
\eear
We observe that, because the gauge fixing conditions we used here are similar to the ones in the preceding paper \cite{Melichev:2023lwj}, we obtained similar expressions for the ghost operator despite having a different gauge group, compared to the Poincar\'e gauge theory.
However, owing to the fact that the volume of the gauge group is different, one will get a different contribution to the path integral.

%%%%%%%%%%%%%%%%%%%%%%%%%%%%%%%%
\section{\label{sec:beta_functions}One loop divergences and beta functions}
%%%%%%%%%%%%%%%%%%%%%%%%%%%%%%%%

\subsection{One loop divergences}

The one-loop effective action is given by a sum of the classical action and quantum contributions
\be \label{eff_action_sum_qcl}
\G=S+\d\G \, ,
\ee
where $\d\G$ can be seen as a perturbative series in $\hbar$.
Considering the UV behavior, we disregard non-local terms and focus in particular on the logarithmically divergent part, which in the presence of a momentum cutoff $\Lambda$ can be written as
\be \label{gamma_log_part}
\d\G^{(1)}_{log} = - \dot{\G}\log\left(\frac{\Lambda}{\m}\right) \, .
\ee
Here $\m$ is a reference scale that has to be introduced for dimensional reasons and the dot denotes derivative over $\log \m$.
The functional $\dot{\G}$ can be seen as the ``beta functional" and presented as a sum of local operators constructed with the fields and their covariant derivatives:
\be \label{beta_functional_definition}
\dot{\G} = \sum_i \int d^4x\sqrt{g}\,\b_i\cL_i \, .
\ee
The bases for the operators $\cL_i$ are the same as for the classical Lagrangian and were presented in section \ref{sec:general_Lag}, while the coefficients $\b_i$ are now the beta functions defined by this equation.
For our theory, the term linear in $\hbar$ in \eqref{eff_action_sum_qcl} is
\be
\d\G^{(1)}=\frac12\Tr\log \D_{g\theta\! A}-\Tr\log\D_{gh} \, . 
\ee
where both $\D_{g\theta\! A}$ and $\D_{gh}$ are operators
of the form 
\be
-\bar\nabla^2+\mathbb{V}^\m\bar\nabla_\m+\mathbb{W} \, , 
\ee
Then the ``beta functional" is
\be \label{beta_functional_with_ghosts}
\dot{\G} = -\Lambda\frac{\partial\G^{(1)}_{log}}{\partial\Lambda} = B_4 \left( \D_{g\theta\! A} \right) - 2 B_4 \left( \D_{gh} \right) \, .
\ee
Defining $\nabla'=\nabla+\frac12 \mathbb{V}$ and redefining $\mathbb{W}$,
the term with one derivative can be removed.
Note that the connection $\nabla'$ now has vectorial torsion and nonmetricity.
However, one can still apply the standard formulae for the Seeley--DeWitt coefficients, which give the following logarithmically divergent part of $\Tr\log\D_{g\theta\! A}$ (and analogously for the ghost operator):
\bear \label{logdiv}
B \left( \D_{g\theta\! A} \right)
= &-\frac{1}{16\pi^2}\int d^4 x\sqrt{g}\,\Tr\Big\{
\frac{1}{180}\left(\bar R_{\m\n\r\sigma}\bar R^{\m\n\r\sigma}
-\bar R_{\m\n}\bar R^{\m\n}+\frac52 \bar R^2\right) \mathbb{1}
\\
&
+\frac12\mathbb{W}^2
-\frac12 \mathbb{W}\, \bar\nabla_\m \mathbb{V}^\m
+\frac14\, \mathbb{W}\, \mathbb{V}_\m \mathbb{V}^\m
-\frac16\bar R\, \mathbb{W}
+\frac{1}{12}\bar R\,\bar\nabla_\m \mathbb{V}^\m
-\frac{1}{24}\bar R\, \mathbb{V}_\m \mathbb{V}^\m
\\&
+\frac{1}{12}\Omega_{\m\n}\Omega^{\m\n}
-\frac16\,\Omega_{\m\n}\bar\nabla^\m \mathbb{V}^\n
+\frac{1}{24}\,\Omega_{\m\n}[\mathbb{V}^\m, \mathbb{V}^\n]
+\frac{1}{8}\bar\nabla_\m \mathbb{V}^\m\bar\nabla_\r \mathbb{V}^\r
\\
&
-\frac{1}{8} \mathbb{V}_\m \mathbb{V}^\m \bar \nabla_\r \mathbb{V}^\r
+\frac{1}{32} \mathbb{V}_\m \mathbb{V}^\m \mathbb{V}_\r \mathbb{V}^\r 
+\frac{1}{24}(\bar\nabla_\m \mathbb{V}_\n-\bar\nabla_\n \mathbb{V}_\m)\bar\nabla^\m \mathbb{V}^\n
\\
&
-\frac{1}{24}\,\bar\nabla_\m \mathbb{V}_\n[\mathbb{V}^\m,\mathbb{V}^\n]
+\frac{1}{192}\,[\mathbb{V}_\m,\mathbb{V}_\n][\mathbb{V}^\m,\mathbb{V}^\n]
\Big\} \, . 
\eear 
This formula can also be obtained directly from the expansion of $\Tr\log$ and the application of the off-diagonal heat kernel techniques
\cite{Barvinsky:1985an, Groh:2011dw, pedagogical}.

%%%%%%%%%
\subsection{\label{subsec:results}Off-shell Results}
%%%%%%%%%

Summing $B_{g\theta\! A}-2B_{gh}$ in \eqref{beta_functional_with_ghosts} from \eqref{B_XYZ} and \eqref{b4DeltaGhostEinsteinRPhi}, and removing the linearly dependent operators,
we arrive at the expression of the divergent terms in the Cartan form.
We present here the marginal (dimension four) contributions in the case of the vanishing Palatini term, while the general expression can be found in the appendix \ref{sec:app:final_result} formula \eqref{Gammadot_final_Cartan}. 
\begingroup
\allowdisplaybreaks
\begin{align*}  \label{Gammadot_final} 
\dot{\G} = & - \frac{1}{32 \pi^2} \int d^4 x \sqrt{g} 
\left[
\left(\frac{37}{6} 
- \frac{a_4}{3 a_1}\right) L^{FF}_{1} 
+ \left(\frac{3}{20} 
- \frac{a_4}{3 a_1}\right) L^{FF}_{2} 
+ \frac{1177}{360} L^{FF}_{3} 
\right. \\ & 
+ \frac{527}{60} L^{FF}_{4} 
- \frac{1061}{60} L^{FF}_{5} 
- \frac{1297}{120} L^{FF}_{6} 
- \frac{33}{5} L^{FF}_{7} 
+ \frac{3797}{360} L^{FF}_{8} 
+ \frac{551}{60} L^{FF}_{9} 
\\ &  \numberthis 
+ \frac{184}{45} L^{FF}_{10} 
- \frac{41}{60} L^{FF}_{11}
+ \frac{478}{45} L^{FF}_{12} 
+ \frac{47}{15} L^{FF}_{13} 
+ \frac{199}{30} L^{FF}_{14} 
- \frac{171}{10} L^{FF}_{15} 
\\ &  
- \frac{119}{36} L^{FF}_{16} 
- 11 L^{FT}_{1} 
- \frac{117}{10} L^{FT}_{10}
+ \frac{122}{5} L^{FT}_{11} 
+ \frac{193}{15} L^{FT}_{12} 
+ \frac{2}{5} L^{FT}_{13} 
- \frac{853}{30} L^{FT}_{14} 
\\ & 
+ \frac{147}{10} L^{FT}_{15} 
+ \frac{93}{10} L^{FT}_{17} 
- \frac{181}{30} L^{FT}_{21}
+ \frac{207}{20} L^{FQ}_{10} 
+ \frac{17}{60} L^{FQ}_{11} 
+ \frac{331}{60} L^{FQ}_{12} 
+ \frac{97}{15} L^{FQ}_{14} 
\\ & 
+ \frac{37}{3} L^{FQ}_{16} 
- \frac{188}{15} L^{FQ}_{17} 
- \frac{193}{20} L^{FQ}_{18} 
- \frac{87}{20} L^{FQ}_{19} 
+ \frac{83}{30} L^{FQ}_{23} 
+ \frac{31}{120} L^{TT}_{1} 
- \frac{139}{20} L^{TT}_{2} 
\\ &
+ \frac{109}{15} L^{TT}_{3} 
+ \frac{122}{15} L^{TT}_{5} 
- \frac{28}{15} L^{TQ}_{1} 
+ \frac{13}{15} L^{TQ}_{10} 
- \frac{73}{10} L^{TQ}_{11} 
- \frac{28}{5} L^{TQ}_{12} 
\\ & \left.
+ \left(\frac{71}{120} + \frac{a_4}{4 a_1}\right) 
L^{QQ}_{1} 
- \frac{11}{60} L^{QQ}_{10} 
- \frac{14}{5} L^{QQ}_{11} 
+ \frac{17}{10} L^{QQ}_{12} 
+ \frac{173}{120} L^{QQ}_{14} 
+ \dots \right] \, 
\end{align*}
\endgroup
The ellipses stand for higher powers of $F$, $T$, $Q$ that do not contribute to the 2-point function around flat space.
This is the main result of the paper.

Let us see now how we can obtain the expression for the beta functional generated by the action \eqref{lag_Donoghue}.
Looking at the $R-\dis$ Einstein representation \eqref{lag_XYZ_Einstein} of the larger action \eqref{lag_XYZ_ours} we observe, that the role of all the $a$ coefficients is merely to give masses to the distortion (alternatively, in \eqref{Lag_XYZ_EinsteinRTQ} they give masses to the torsion and the nonmetricity fields).
This means that when we change these coefficients, up to the limit $a_0,a_1,a_4\rightarrow0$, the gauge group remains the same.
Considered around Minkowski space, several additional states may be propagating, on top of the graviton.
The introduction of the mass terms cannot add new ones to the spectrum, and neither it can affect the ultraviolet properties of the theory.
Assuming there is no unexpected discontinuity in the massless limit, the divergences generated by the action \eqref{lag_Donoghue} can be recovered from \eqref{Gammadot_final} by taking the limit $a_4 \rightarrow 0$.
In the next section, we will further comment on extracting a universal result from it.

%%%%%%%%%%%%%%%%%%%%%%%%%%%
\section{\label{sec:on_shell}On-shell reduction of the effective action}

In the previous sections, we have worked out one-loop counterterms without considering the equations of motion.
We will see now, that the result \eqref{Gammadot_final} can be simplified on the mass shell under certain conditions.
As previously, we work in the perturbative regime, which means, in particular, that 
\be \label{one_loop_perturbative_condition}
S \gg \hbar\, \d \Gamma^{(1)} \, .
\ee
Let us consider the following infinitesimal redefinitions of the set of dynamical fields $\psi = \left(A, \theta, g \right)^T$:
\be \label{field_redefs_general}
\psi \rightarrow \psi + \Psi[\psi],
\ee
where $ \psi \gg \Psi[\psi]$.
The corresponding change of the effective action is
\be
\label{Gamma_field_redef_change}
\G[\psi] \rightarrow \G[\psi] + \frac{\d \G}{\d \psi}\Psi [\psi].
\ee
This means that terms proportional to the equations of motion can be (infinitesimally) eliminated from the effective action and, at the one-loop level, we can define the on-shell effective action as:
\be
\label{on-shell_reduction_1-loop}
\d\G^{(1)} \approx \d\G^{(1)}_{on-shell} + \frac{\d S}{\d \psi}\Psi [\psi].
\ee

Starting from the action \eqref{lag_XYZ_ours} we observe that the presence of the mass terms for torsion and nonmetricity suppresses them at low energies.
This means that all terms involving $T$, $Q$ or $\dis$
are on-shell redundant.
The analog of that in GR is the vanishing of the Ricci tensor on shell and the redundancy of the $R^2$ and $R_{\m\n}R^{\m\n}$ terms.
However, looking at the formula \eqref{logdiv} we notice that $\mathbb{W}$ contains quadratic powers of curvature, and the term $1/2\, \Tr\, \mathbb{W}^2$, as well as some other terms, will produce forth powers of Riemann curvature, and such terms are essential.
This implies the nonrenormalizability of the theory at one loop, similar to Poincar\'e gauge theory \cite{Melichev:2023lwj}.

Consider now the theory presented by the action \eqref{lag_Donoghue}, which can be presented in the Einstein form as
\bear \label{lag_Donoghue_Einstein}
S_{YM} = - c_1 \int d^4 x \sqrt{g}\, \left[ H^{RR}_{1} + 4 H^{R\dis}_1 + 2 H^{\dis\dis}_{1} - 2 H^{\dis\dis}_{12} \right] \, .
\eear
Here there is no mass term and no low-energy suppression of the dynamics of torsion and nonmetricity happens.
One can, however, still somewhat reduce the number of terms generated at one loop.
To this end, we consider the equations of motion in the Cartan form.
The variation of the action \eqref{lag_Donoghue} over the metric is trivial, while variations of the soldering and connection give
\begin{subequations}
\begin{align}
\label{eom_Donoghue_X}
\frac{\d S_{YM}}{\d \theta^\a{}_\b} &= c_1 \left[ L^{FF}_1 \d^\a_\b - 4 F^{\a\m\n\r} F_{\b\m\n\r} \right] \, ,
\\
\label{eom_Donoghue_Z}\frac{\d S_{YM}}{\d A_\a{}^\b{}_\g} &= 4 c_1 \left[ \na_\r F^{\a\r}{}_\b{}^\g + F^\a{}_{\r\b\l} \dis^{\r\g\l}
- F^\a{}_{\r\l}{}^\g \dis^{\r\l\b}
\right] \, .
\end{align}
\end{subequations}
The former equation does not yield any nontrivial relations between the dimension four terms, while the latter does.
Contracting it with the distortion tensor in 15 different ways gives 15 independent relations that are presented in the appendix \ref{sec:app:eoms_Donoghue}.
We can use these relations to simplify the effective action on shell.
This is equivalent to performing linear perturbative field redefinitions of the type \eqref{field_redefs_general}, namely,
\be
A_\m{}^\r{}_\s \rightarrow A_\m{}^\r{}_\s + \hbar\, B_\m{}^\r{}_{\s,\,}{}^\a{}_\b{}^\g \, A_\a{}^\b{}_\g \, ,
\ee
where $B$ contains merely Kronecker deltas and coupling constants and produces 15 possible permutations of Lorentz indices.
Even more than for the off-shell action, there exists, of course, ambiguity when it comes to choosing a basis of on-shell independent terms.
We will favor the curvature-squared contributions \eqref{LFF} as they are the ones more commonly used in literature.
Looking at our expression for the on-shell divergence \eqref{Gammadot_final} we observe that in the limit $a_4\rightarrow 0$ there are 47 independent terms with numerical coefficients.
The aforementioned relations allow us to reduce the number of terms to 32, producing the following one-loop on-shell divergence:
\bear \label{Gammadot_YangMills_onshell}
\dot{\G}&_{YM}^{on-shell} = - \frac{1}{32 \pi^2} \int d^4 x \sqrt{g} 
\left[
\frac{769}{30} L^{FF}_{1} - \frac{3377}{360} L^{FF}_{3} + \frac{69}{40} L^{FF}_{6} + \frac{325}{72} L^{FF}_{8} - \frac{67}{20} L^{FF}_{9} 
\right. \\ & \qquad
+ \frac{184}{45} L^{FF}_{10} - \frac{73}{36} L^{FF}_{12} 
+ \frac{493}{60} L^{FF}_{15} - \frac{119}{36} L^{FF}_{16} + \frac{343}{60} L^{FT}_{11} + \frac{261}{20} L^{FT}_{13} + \frac{54}{5} L^{FT}_{14} 
\\ & \qquad
+ \frac{41}{20} L^{FT}_{15} - \frac{563}{60} L^{FT}_{17} + \frac{17}{60} L^{FQ}_{11} - \frac{91}{10} L^{FQ}_{16} + \frac{673}{60} L^{FQ}_{18} - \frac{87}{20} L^{FQ}_{19} - \frac{107}{30} L^{FQ}_{23} 
\\ & \qquad
+ \frac{31}{120} L^{TT}_{1} - \frac{139}{20} L^{TT}_{2} + \frac{109}{15} L^{TT}_{3} + \frac{122}{15} L^{TT}_{5} - \frac{28}{15} L^{TQ}_{1} + \frac{13}{15} L^{TQ}_{10} - \frac{73}{10} L^{TQ}_{11} 
\\ & \qquad \left.
- \frac{28}{5} L^{TQ}_{12} + \frac{71}{120} L^{QQ}_{1} - \frac{11}{60} L^{QQ}_{10} - \frac{14}{5} L^{QQ}_{11} + \frac{17}{10} L^{QQ}_{12} + \frac{173}{120} L^{QQ}_{14} 
+ \dots \right] \, .
\eear
The on-shell results are known to be gauge-independent and now it makes sense to make a comparison with the results presented in the literature.
The result of \cite{Donoghue:2016vck} suggests that the beta function of the Yang--Mills coupling is
\be \label{beta_function_c1_result_Donoghue}
\b (g) = - \frac{g^3}{16 \pi^2} \frac{22}{3} \, .
\ee
Let us look at the first term of our result \eqref{Gammadot_YangMills_onshell}, which exactly corresponds to the running of the Yang--Mills-like term present in the original Lagrangian \eqref{lag_Donoghue}.
Comparing it with the two expressions we read off the beta function as
\be
\b (c_1) = \frac{1}{16 \pi^2} \frac{769}{30} \, ,
\ee
and noticing that the standard prefactor is $-1/4g^2$ instead of $-c_1/2$ we obtain for the beta function of $g$ as
\be \label{beta_function_c1_result_mine}
\b (g) = - \frac{g^3}{16 \pi^2} \frac{769}{30} \, .
\ee
We see, that it has a negative sign, which suggests that, qualitatively, conclusions of \cite{Donoghue:2016vck} regarding the asymptotic freedom of this theory may be correct.
Quantitatively, however, we do not see an agreement, and more importantly, we see that many more terms are generated, which are just as important as the first Yang--Mills-like term.
Let us also remind here that formulae \eqref{Gammadot_final}, \eqref{Gammadot_final_Cartan}, and \eqref{Gammadot_YangMills_onshell} do not display the terms that contain higher powers of $F$, $T$, $Q$, which are of the same order in mass dimension as the ones that are displayed, but can only contribute to interactions and not to the 2-point function around flat space.
This means, that in order to have a perturbatively renormalizable theory and identify the UV critical surface connected to free fixed points one has to introduce other terms into the original Lagrangian.

%%%%%%%%%%%%%%%%%%%%%%%%%%%%%%%%%%%%%%%%%%
\section{\label{sec:Discussion}Discussion}
%%%%%%%%%%%%%%%%%%%%%%%%%%%%%%%%%%%%%%%%%%

We presented a detailed computation of the logarithmically divergent part of the effective action at one loop, generated by the action \eqref{lag_XYZ_ours}.
Following the standard Faddeev-Popov procedure, we accounted for the gauge freedom by considering the ghost contribution.
A na\"ive calculation of the beta function by considering the analogy between MAG and the standard Yang--Mills theory for the gauge group $SU(N)$ had been previously performed in \cite{Donoghue:2016vck}.
That result was quoted in \cite{Alexander:2022acv}, and rederived in \cite{Boos:2023xoq}, with the same considerations repeated.
As mentioned in the introduction, it does not agree with those we obtain even on shell.
Let us understand why is that the case.

Firstly, as discussed in section \ref{sec:mag_as_gauge_theory}, in vierbein formulation used in \cite{Donoghue:2016vck}, the gauge group is not $O(1,3)$ but 
$\mathbb{R}^{1, 3} \rtimes O(1,3)$, which means that the ghost contribution will differ from the one appearing in the standard Yang--Mills theory.
Secondly, the computation of \cite{Donoghue:2016vck} only takes into account the $\d A-\d A$ ($Z-Z$) sector of the kinetic operator responsible for the dynamics of the connection.
In fact, there exists also the $\d\theta-\d\theta$ ($X-X$) sector, as well as kinetic mixing and contributions from the determinant of the soldering (vierbein).
That is why it is incorrect to extract the results for MAG using the standard results for the Yang--Mills theory by just replacing the Casimir operator in the way that it was done in \cite{Donoghue:2016vck}.
Ironically, by analogous consideration applied to the Poincar\'e gauge theory, one could conclude that the beta function of the same coupling is zero (\cite{Boos:2023xoq}, section IV), which certainly contradicts with what was previously computed directly in \cite{Melichev:2023lwj}.
Lastly, as has been reiterated throughout the paper, RG running does generate many other terms allowed by symmetries.
In some cases, different terms of the same dimension can be absorbed into each other by linear or nonlinear field redefinitions.
Therefore, it makes little sense to compute the running of a single individual term such as the one present in the original Lagrangian, unless an explanation of why its running has independent physical significance is provided.

The calculations presented in this paper, as well as possible generalizations, face problems pervasive in quantum field theoretic description of gravity.
Leaving aside the general discussion of the validity of the background field split, after which the metric and its perturbations are treated in different ways, we only comment on the two issues that do appear in metric gravity but that become more apparent in the context of MAG.
As discussed in the previous section, the existence of the possibility to perform field redefinitions tells us that the beta functions (as well as any terms in the effective action, infinite of finite) depend on the way we \emph{define} the RG flow.
By choosing different $\Psi$
in \eqref{field_redefs_general}, one can alter the RG trajectory.
It is similar to what happens in any quantum field theory, and the context of (nonperturbative) quantization of metric gravity this is known as the essential renormalization group \cite{Pawlowski:2005xe, Anselmi:2002ge, Baldazzi:2021ydj, Baldazzi:2021orb, Baldazzi:2023pep}.
Physical theory corresponds to an equivalence class of RG trajectories rather than a single trajectory.
In MAG we have many more terms than in metric gravity, which translates to a greater freedom to choose renormalization conditions.

The other problem that is exacerbated in MAG is the complexity of the kinetic operators and mixing of degrees of freedom.
In the relatively simple setup discussed in this paper, we utilized the possibility of making the kinetic operator minimal by an appropriate gauge choice.
The general space of MAG theories consists of those yielding nonminimal kinetic operators, apart from a measure zero subspace.
The way of dealing with nonminimal operators exists in principle \cite{Barvinsky:1985an}, but its application often leads to involved computations.
If we try to apply the method used in section \ref{sec:loop_calculation} to a theory with an action containing $(DT)^2$ term, we arrive at the operator with a nonminimal principal part.
In some cases, operators with a degenerate principal part appear\footnote{A kinetic operator is said to have a degenerate principal (= highest order in derivatives) part if it is not invertible while the full operator is.}.
Such a degeneracy can also be gauge-dependent, in the sense that the same theory in different variables %(such as, for example, $\dis, g$ Einstein and $A, g$ Cartan forms) 
produce different kinetic operators, one of which has a degenerate principal part and one does not.
We conclude that if a general study of loop corrections in MAG is to be conducted, efficient techniques and/or software able to deal with higher-rank nonminimal Laplace-type operators must be developed.

%%%%%%%%%%%%%%%%%%%%
\section{\label{sec:Acknowledgments}Acknowledgments}

The author is indebted to Roberto Percacci and Alexander Ochirov for useful suggestions and comments upon reading an earlier draft of the paper.
The {\tt xAct} bundle of packages, in particular, {\tt xTensor} \cite{xAct:xTensor}, {\tt Invar} \cite{Martin-Garcia:2007bqa, Martin-Garcia:2008yei}, {\tt SymManipulator} \cite{xAct:SymManipulator}, and {\tt xTras} \cite{Nutma:2013zea} were extensively used to perform the computations.

\begin{appendix}

%%%%%%%%%%%%%%%%%%%%%%
\section{\label{sec:app:final_result}Final Off-shell result}

Here we present the final expression for the logarithmic divergence generated by the action \eqref{lag_XYZ_ours} in the Cartan form:
\begingroup
\small
\allowdisplaybreaks
\begin{align*} 
\dot{\G} &= - \frac{1}{32 \pi^2} \int d^4 x \sqrt{g} \left[
 \left(\frac{37}{6} + \frac{17 a_0^2}{48 a_1^2} + \frac{29 a_0}{24 a_1} - \frac{a_4}{3 a_1}\right) L^{FF}_{1} + \left(\frac{3}{20} - \frac{53 a_0^2}{48 a_1^2} + \frac{5 a_0}{4 a_1} - \frac{a_4}{3 a_1}\right) L^{FF}_{2} 
\right.
\\ & 
+ \left(\frac{1177}{360} + \frac{7 a_0^2}{12 a_1^2} + \frac{a_0}{6 a_1}\right) L^{FF}_{3} 
+ \left(\frac{527}{60} + \frac{a_0^2}{6 a_1^2} - \frac{15 a_0}{8 a_1}\right) L^{FF}_{4} 
+ \left(- \frac{1061}{60} - \frac{a_0^2}{3 a_1^2} + \frac{29 a_0}{12 a_1}\right) L^{FF}_{5} 
\\ & 
+ \left(- \frac{1297}{120} + \frac{a_0^2}{6 a_1^2} - \frac{59 a_0}{24 a_1}\right) L^{FF}_{6} + \left(- \frac{33}{5} + \frac{5 a_0^2}{12 a_1^2} + \frac{9 a_0}{8 a_1}\right) L^{FF}_{7} 
+ \left(\frac{3797}{360} + \frac{a_0^2}{24 a_1^2} + \frac{7 a_0}{8 a_1}\right) L^{FF}_{8} 
\\ & 
+ \left(\frac{551}{60} + \frac{5 a_0^2}{12 a_1^2} + \frac{15 a_0}{8 a_1}\right) L^{FF}_{9} + \left(\frac{184}{45} + \frac{a_0^2}{24 a_1^2} - \frac{a_0}{24 a_1}\right) L^{FF}_{10} + \left(- \frac{41}{60} - \frac{5 a_0^2}{6 a_1^2} - \frac{5 a_0}{4 a_1}\right) L^{FF}_{11} \\ & 
+ \left(\frac{478}{45} - \frac{a_0^2}{12 a_1^2} + \frac{3 a_0}{4 a_1}\right) L^{FF}_{12} + \frac{47}{15} L^{FF}_{13} + \left(\frac{199}{30} + \frac{a_0}{a_1}\right) L^{FF}_{14} + \left(- \frac{171}{10} - \frac{11 a_0}{4 a_1}\right) L^{FF}_{15} 
\\ & 
+ \left(- \frac{119}{36} + \frac{a_0^2}{12 a_1^2} - \frac{5 a_0}{12 a_1}\right) L^{FF}_{16} + \left(-11 - \frac{23 a_0}{12 a_1}\right) L^{FT}_{1} + \left(- \frac{117}{10} - \frac{7 a_0}{4 a_1}\right) L^{FT}_{10} 
\\ & 
+ \left(\frac{122}{5} + \frac{7 a_0}{4 a_1}\right) L^{FT}_{11} + \frac{193}{15} L^{FT}_{12} + \left(\frac{2}{5} - \frac{7 a_0}{12 a_1}\right) L^{FT}_{13} + \left(- \frac{853}{30} - \frac{23 a_0}{12 a_1}\right) L^{FT}_{14} 
\\ & 
+ \left(\frac{147}{10} 
+ \frac{7 a_0}{12 a_1}\right) L^{FT}_{15} 
+ \left(\frac{93}{10} + \frac{7 a_0}{4 a_1}\right) L^{FT}_{17} + \left(- \frac{181}{30} - \frac{a_0}{2 a_1}\right) L^{FT}_{21} + \left(\frac{207}{20} + \frac{7 a_0}{4 a_1}\right) L^{FQ}_{10} 
\\ & 
+ \left(\frac{17}{60} - \frac{a_0}{12 a_1}\right) L^{FQ}_{11} + \frac{331}{60} L^{FQ}_{12} 
+ \frac{97}{15} L^{FQ}_{14} 
+ \left(\frac{37}{3} + \frac{5 a_0}{4 a_1}\right) L^{FQ}_{16} + \left(- \frac{188}{15} - \frac{5 a_0}{4 a_1}\right) L^{FQ}_{17} 
\\ & 
+ \left(- \frac{193}{20} - \frac{7 a_0}{4 a_1}\right) L^{FQ}_{18} 
+ \left(- \frac{87}{20} + \frac{a_0}{12 a_1}\right) L^{FQ}_{19} 
+ \left(\frac{83}{30} + \frac{5 a_0}{12 a_1}\right) L^{FQ}_{23} 
\\ & 
+ \frac{31}{120} L^{TT}_{1} - \frac{139}{20} L^{TT}_{2} 
+ \frac{109}{15} L^{TT}_{3} 
+ \frac{122}{15} L^{TT}_{5} - \frac{28}{15} L^{TQ}_{1} 
+ \frac{13}{15} L^{TQ}_{10} 
- \frac{73}{10} L^{TQ}_{11} 
\\ & 
- \frac{28}{5} L^{TQ}_{12} + \left(\frac{71}{120} + \frac{a_4}{4 a_1}\right) L^{QQ}_{1} - \frac{11}{60} L^{QQ}_{10} 
- \frac{14}{5} L^{QQ}_{11} + \frac{17}{10} L^{QQ}_{12} + \frac{173}{120} L^{QQ}_{14} 
\label{Gammadot_final_Cartan} \numberthis 
\\ & 
+ \left(\frac{17 a_0}{8 c_1} + \frac{371 a_0^2}{96 a_1 c_1} - \frac{347 a_1}{24 c_1}\right) m^{TT}_{1} + \left(\frac{2 a_0}{c_1} + \frac{109 a_0^2}{48 a_1 c_1} - \frac{175 a_1}{12 c_1}\right) m^{TT}_{2} 
\\ & 
+ \left(- \frac{21 a_0}{4 c_1} - \frac{35 a_0^2}{12 a_1 c_1} + \frac{35 a_1}{3 c_1}\right) m^{TT}_{3} + \left(- \frac{41 a_0}{8 c_1} - \frac{55 a_0^2}{8 a_1 c_1} + \frac{97 a_1}{2 c_1}\right) m^{TQ}_{1} 
\\ & 
+ \left(- \frac{39 a_0}{8 c_1} - \frac{41 a_0^2}{24 a_1 c_1} + \frac{163 a_1}{12 c_1}\right) m^{TQ}_{2} + \left(\frac{35 a_0}{8 c_1} + \frac{41 a_0^2}{24 a_1 c_1} - \frac{37 a_1}{3 c_1}\right) m^{TQ}_{3} 
\\ & 
+ \left(\frac{a_0}{2 c_1} + \frac{21 a_0^2}{32 a_1 c_1} - \frac{53 a_1}{8 c_1} + \frac{5 a_4}{2 c_1} + \frac{3 a_0 a_4}{4 a_1 c_1}\right) m^{QQ}_{1} + \left(- \frac{a_0}{8 c_1} - \frac{2 a_0^2}{a_1 c_1} + \frac{12 a_1}{c_1} + \frac{a_4}{c_1} + \frac{a_0 a_4}{a_1 c_1}\right) m^{QQ}_{2} 
\\ & 
+ \left(- \frac{7 a_0}{8 c_1} - \frac{a_0^2}{12 a_1 c_1} + \frac{7 a_1}{3 c_1} - \frac{a_4}{c_1}\right) m^{QQ}_{3} + \left(- \frac{7 a_0}{8 c_1} - \frac{7 a_0^2}{48 a_1 c_1} + \frac{31 a_1}{12 c_1} - \frac{a_4}{c_1} - \frac{a_0 a_4}{a_1 c_1}\right) m^{QQ}_{4} 
\\ & 
\left.
+ \left(\frac{11 a_0}{8 c_1} + \frac{35 a_0^2}{48 a_1 c_1} - \frac{83 a_1}{12 c_1}\right) m^{QQ}_{5} + \left(\frac{a_0}{4 c_1} - \frac{19 a_0^3}{16 a_1^2 c_1} + \frac{13 a_0^2}{8 a_1 c_1} + \frac{5 a_1}{2 c_1}\right) F
+ \frac{9 a_0^4}{8 a_1^2 c_1^2} + \frac{18 a_1^2}{c_1^2} + \dots \right] \, .
\end{align*} 
\endgroup
%\eqref{Gammadot_final_Cartan}
Here we expressed it in the following basis of the Cartan form that is used throughout the paper:
\bear
\label{cargenbasisFF}
&\{ L^{FF}_1\, , \dots , \ L^{FF}_{16} \}\\
&\{ L^{TT}_1\, , \ L^{TT}_2\, , \ 
L^{TT}_3\, , \ 
L^{TT}_5 \}\\ 
&\{L^{QQ}_1 \, , \ L^{QQ}_{10} \, , \ L^{QQ}_{11}\, , \ L^{QQ}_{12} \, , \ L^{QQ}_{14} \} \\ 
& \{ L^{TQ}_1, L^{TQ}_{10}, L^{TQ}_{11}, L^{TQ}_{12} \}\\ 
&\{L^{FT}_1\, , \ L^{FT}_{10}\, , \ L^{FT}_{11} , \ L^{FT}_{12}\, , \ L^{FT}_{13}\, , \ L^{FT}_{14}\, , \ L^{FT}_{15}\, , \ L^{FT}_{17}\, , \ L^{FT}_{21}\}\\ 
&\{ L^{FQ}_{10} \, , \ L^{FQ}_{11} \, , \ L^{FQ}_{12} \, , \ L^{FQ}_{14} \, , \ L^{FQ}_{16} \, , \ L^{FQ}_{17} \, , \ L^{FQ}_{18} \, , \ L^{FQ}_{19}\, , \ L^{FQ}_{23} \} \ .
\eear
We refer to \cite{Baldazzi:2021kaf}, section 3 for a complete listing of all the other contributing to the flat space 2-point function invariants.
The same expression in the Einstein form is somewhat longer:
\begingroup
\small
\allowdisplaybreaks
\begin{align*} 
\dot{\G} &= - \frac{1}{32 \pi^2} \int d^4 x \sqrt{g} \left[
\left(\frac{12323}{720} + \frac{19 a_0^2}{8 a_1^2} - \frac{13 a_0}{4 a_1}\right) H^{RR}_{1} 
+ \left(\frac{2621}{360} + \frac{11 a_0^2}{6 a_1^2} + \frac{13 a_0}{3 a_1}\right) H^{RR}_{2} 
\right.
\\ & 
+ \left(- \frac{119}{36} + \frac{a_0^2}{12 a_1^2} - \frac{5 a_0}{12 a_1}\right) H^{RR}_{3} 
+ \left(\frac{391}{4} + \frac{68 a_0^2}{3 a_1^2} - \frac{67 a_0}{3 a_1}\right) H^{RT}_{3} 
+ \left(- \frac{45}{8} + \frac{13 a_0^2}{6 a_1^2} + \frac{13 a_0}{6 a_1}\right) H^{RT}_{5} 
\\ & 
+ \left(- \frac{407}{8} - \frac{34 a_0^2}{3 a_1^2} + \frac{67 a_0}{6 a_1}\right) H^{RQ}_{4} 
+ \left(\frac{629}{12} + \frac{34 a_0^2}{3 a_1^2} - \frac{61 a_0}{6 a_1}\right) H^{RQ}_{5} 
+ \left(\frac{143}{48} - \frac{13 a_0^2}{12 a_1^2} - \frac{11 a_0}{12 a_1}\right) H^{RQ}_{6} 
\\ & 
+ \left(- \frac{29}{8} + \frac{13 a_0^2}{12 a_1^2} + \frac{5 a_0}{12 a_1}\right) H^{RQ}_{7} 
+ \left(\frac{739}{96} + \frac{113 a_0^2}{48 a_1^2} - \frac{65 a_0}{24 a_1}\right) H^{TT}_{1} 
+ \left(\frac{331}{48} + \frac{43 a_0^2}{24 a_1^2} - \frac{41 a_0}{12 a_1}\right) H^{TT}_{2} 
\\ &
+ \left(- \frac{7}{6} + \frac{5 a_0^2}{3 a_1^2} + \frac{5 a_0}{2 a_1}\right) H^{TT}_{3} 
+ \left(- \frac{31}{8} + \frac{23 a_0^2}{24 a_1^2} - \frac{17 a_0}{12 a_1}\right) H^{TT}_{4} 
+ \left(\frac{127}{24} + \frac{31 a_0^2}{8 a_1^2} - \frac{41 a_0}{12 a_1}\right) H^{TT}_{5} 
\\ & 
+ \left(- \frac{275}{48} - \frac{53 a_0^2}{48 a_1^2} + \frac{83 a_0}{24 a_1}\right) H^{TT}_{6} 
+ \left(- \frac{121}{8} - \frac{43 a_0^2}{12 a_1^2} + \frac{49 a_0}{6 a_1}\right) H^{TT}_{7} 
\label{Gammadot_final_Einstein} \numberthis
\\ &
+ \left(\frac{61}{6} - \frac{10 a_0^2}{3 a_1^2} - \frac{22 a_0}{3 a_1}\right) H^{TT}_{8} 
+ \left(\frac{25}{8} + \frac{a_0^2}{2 a_1^2} - \frac{5 a_0}{6 a_1}\right) H^{TT}_{9} 
\\ & 
+ \left(-22 - \frac{13 a_0^2}{2 a_1^2} + \frac{15 a_0}{2 a_1}\right) H^{TQ}_{1} 
+ \left(3 - \frac{5 a_0^2}{3 a_1^2} - \frac{3 a_0}{2 a_1}\right) H^{TQ}_{2} 
\\ & 
+ \left(- \frac{5}{2} + \frac{5 a_0^2}{3 a_1^2} + \frac{3 a_0}{2 a_1}\right) H^{TQ}_{3} + \left(- \frac{31}{8} - \frac{29 a_0^2}{6 a_1^2} + \frac{37 a_0}{6 a_1}\right) H^{TQ}_{4} + \left(\frac{125}{6} + \frac{4 a_0^2}{a_1^2} - \frac{31 a_0}{3 a_1}\right) H^{TQ}_{5} 
\\ &
+ \left(\frac{337}{24} + \frac{17 a_0^2}{3 a_1^2} - \frac{7 a_0}{a_1}\right) H^{TQ}_{6} + \left(- \frac{235}{24} - \frac{5 a_0^2}{6 a_1^2} + \frac{11 a_0}{6 a_1}\right) H^{TQ}_{7} + \left(-6 + \frac{5 a_0^2}{3 a_1^2} + \frac{8 a_0}{3 a_1}\right) H^{TQ}_{8} 
\\ & 
+ \left(\frac{67}{12} - \frac{5 a_0^2}{3 a_1^2} - \frac{8 a_0}{3 a_1}\right) H^{TQ}_{9} 
+ \left(-7 + \frac{5 a_0^2}{3 a_1^2} + \frac{5 a_0}{2 a_1}\right) H^{TQ}_{10} + \left(\frac{43}{8} - \frac{5 a_0^2}{3 a_1^2} - \frac{7 a_0}{2 a_1}\right) H^{TQ}_{11} 
\\ & 
+ \left(- \frac{59}{24} - \frac{a_0^2}{2 a_1^2} + \frac{a_0}{6 a_1}\right) H^{TQ}_{12} 
+ \left(\frac{29}{12} + \frac{a_0^2}{2 a_1^2} - \frac{7 a_0}{6 a_1}\right) H^{TQ}_{13} 
\\ & 
+ 
\left(\frac{373}{96} + \frac{5 a_0^2}{4 a_1^2} - \frac{19 a_0}{24 a_1} - \frac{a_4}{12 a_1}\right) H^{QQ}_{1} 
\\ &
+ \left(- \frac{81}{16} - \frac{13 a_0^2}{8 a_1^2} + \frac{37 a_0}{24 a_1}\right) H^{QQ}_{2} 
+ \left(- \frac{7}{6} + \frac{5 a_0^2}{12 a_1^2} + \frac{a_0}{8 a_1}\right) H^{QQ}_{3} + \left(- \frac{19}{24} + \frac{5 a_0^2}{12 a_1^2} + \frac{a_0}{8 a_1}\right) H^{QQ}_{4} 
\\ & 
+ \left(\frac{23}{12} - \frac{5 a_0^2}{6 a_1^2} - \frac{a_0}{4 a_1}\right) H^{QQ}_{5} 
+ \left(\frac{77}{24} + \frac{19 a_0^2}{12 a_1^2} - \frac{21 a_0}{8 a_1} + \frac{a_4}{3 a_1}\right) H^{QQ}_{6} + \left(- \frac{181}{24} - \frac{29 a_0^2}{24 a_1^2} + \frac{8 a_0}{3 a_1}\right) H^{QQ}_{7} 
\\ & 
+ \left(\frac{229}{24} + \frac{29 a_0^2}{12 a_1^2} - \frac{97 a_0}{24 a_1}\right) H^{QQ}_{8} 
+ \left(- \frac{15}{4} - \frac{29 a_0^2}{12 a_1^2} + \frac{13 a_0}{4 a_1}\right) H^{QQ}_{9} + \left(\frac{13}{3} - \frac{5 a_0^2}{6 a_1^2} - \frac{3 a_0}{4 a_1}\right) H^{QQ}_{10} 
\\ & 
+ \left(- \frac{47}{12} + \frac{5 a_0^2}{6 a_1^2} + \frac{3 a_0}{4 a_1}\right) H^{QQ}_{11} 
+ \left(- \frac{77}{24} + \frac{5 a_0^2}{6 a_1^2} + \frac{5 a_0}{4 a_1}\right) H^{QQ}_{12} + \left(\frac{149}{48} - \frac{5 a_0^2}{6 a_1^2} - \frac{5 a_0}{4 a_1}\right) H^{QQ}_{13} 
\\ & 
+ \left(\frac{71}{96} + \frac{a_0^2}{8 a_1^2} + \frac{a_0}{8 a_1}\right) H^{QQ}_{14} 
+ \left(\frac{1}{2} + \frac{a_0^2}{8 a_1^2} - \frac{3 a_0}{8 a_1}\right) H^{QQ}_{15} + \left(- \frac{25}{24} - \frac{a_0^2}{4 a_1^2} + \frac{a_0}{4 a_1}\right) H^{QQ}_{16} 
\\ &
+ \left(\frac{35 a_0}{16 c_1} - \frac{19 a_0^3}{64 a_1^2 c_1} + \frac{205 a_0^2}{48 a_1 c_1} - \frac{83 a_1}{6 c_1}\right) m^{TT}_{1} 
+ \left(\frac{17 a_0}{8 c_1} - \frac{19 a_0^3}{32 a_1^2 c_1} + \frac{37 a_0^2}{12 a_1 c_1} - \frac{40 a_1}{3 c_1}\right) m^{TT}_{2} 
\\ & 
+ \left(- \frac{11 a_0}{2 c_1} + \frac{19 a_0^3}{16 a_1^2 c_1} - \frac{109 a_0^2}{24 a_1 c_1} + \frac{55 a_1}{6 c_1}\right) m^{TT}_{3} 
+ \left(- \frac{43 a_0}{8 c_1} + \frac{19 a_0^3}{16 a_1^2 c_1} - \frac{17 a_0^2}{2 a_1 c_1} + \frac{46 a_1}{c_1}\right) m^{TQ}_{1} 
\\ & 
+ \left(- \frac{41 a_0}{8 c_1} + \frac{19 a_0^3}{16 a_1^2 c_1} - \frac{10 a_0^2}{3 a_1 c_1} + \frac{133 a_1}{12 c_1}\right) m^{TQ}_{2} + \left(\frac{37 a_0}{8 c_1} - \frac{19 a_0^3}{16 a_1^2 c_1} + \frac{10 a_0^2}{3 a_1 c_1} - \frac{59 a_1}{6 c_1}\right) m^{TQ}_{3} 
\\ & 
+ \left(\frac{9 a_0}{16 c_1} - \frac{19 a_0^3}{64 a_1^2 c_1} + \frac{17 a_0^2}{16 a_1 c_1} - \frac{6 a_1}{c_1} + \frac{5 a_4}{2 c_1} + \frac{3 a_0 a_4}{4 a_1 c_1}\right) m^{QQ}_{1} 
\\ & 
+ \left(- \frac{a_0}{4 c_1} + \frac{19 a_0^3}{32 a_1^2 c_1} - \frac{45 a_0^2}{16 a_1 c_1} + \frac{43 a_1}{4 c_1} + \frac{a_4}{c_1} + \frac{a_0 a_4}{a_1 c_1}\right) m^{QQ}_{2} 
\\ & 
+ \left(- \frac{15 a_0}{16 c_1} + \frac{19 a_0^3}{64 a_1^2 c_1} - \frac{47 a_0^2}{96 a_1 c_1} + \frac{41 a_1}{24 c_1} - \frac{a_4}{c_1}\right) m^{QQ}_{3} 
\\ &
+ \left(- \frac{7 a_0}{8 c_1} - \frac{7 a_0^2}{48 a_1 c_1} + \frac{31 a_1}{12 c_1} - \frac{a_4}{c_1} - \frac{a_0 a_4}{a_1 c_1}\right) m^{QQ}_{4} 
+ \left(\frac{3 a_0}{2 c_1} - \frac{19 a_0^3}{32 a_1^2 c_1} + \frac{37 a_0^2}{24 a_1 c_1} - \frac{17 a_1}{3 c_1}\right) m^{QQ}_{5}
\\ & 
\left.
+ \left(\frac{a_0}{4 c_1} - \frac{19 a_0^3}{16 a_1^2 c_1} + \frac{13 a_0^2}{8 a_1 c_1} + \frac{5 a_1}{2 c_1}\right) R 
+ \frac{9 a_0^4}{8 a_1^2 c_1^2} 
+ \frac{18 a_1^2}{c_1^2} 
+ \dots
\right] \, .
\end{align*} 
\endgroup
%\eqref{Gammadot_final_Einstein}
And for the Einstein form we use the basis 
\bear
\label{EinsteinRTQbasis}
&\{ H^{RR}_1\, , \ H^{RR}_2\, , \ H^{RR}_{3} \}\\
&\{ H^{RT}_3\, , \ H^{RT}_5 \}\\ 
&\{ H^{RQ}_4\, , \ H^{RQ}_5\, , \ H^{RQ}_6\, , \ H^{RQ}_7 \}\\ 
&\{ H^{TT}_1\, , \dots , \ H^{TT}_9 \}\\
&\{ H^{TQ}_1\, , \dots , \ H^{TQ}_{13} \}\\
&\{ H^{QQ}_1\, , \dots , \ H^{QQ}_{16} \} \ .
\eear

%%%%%%%%%%%%%%%%%%%%%%%%%%%
\section{\label{sec:app:bases}Alternative basis}

Certain results can be expressed more compactly in either Einstein ($H^{RR}_i$, $H^{RT}_i$, $H^{RQ}_i$, $H^{TT}_i$, $H^{TQ}_i$, $H^{QQ}_i$) or Cartan ($L^{FF}_i$, $L^{FT}_i$, $L^{FQ}_i$, $L^{TT}_i$, $L^{TQ}_i$, $L^{QQ}_i$) forms.
The third option we use is the basis formulated with the distortion tensor.
We have 38 terms of the $(\nabla\dis)^2$-type:
\be
\begin{tabular}{lll}
$H^{\dis\dis}_{1} = \nabla^\alpha\dis^{\beta\gamma\delta} \nabla_\alpha\dis_{\beta\gamma\delta}$ , 
& $H^{\dis\dis}_{2} = \nabla^\alpha\dis^{\beta\gamma\delta} \nabla_\alpha\dis_{\beta\delta\gamma}$ , 
& $H^{\dis\dis}_{3} = \nabla^\alpha\dis^{\beta\gamma\delta} \nabla_\alpha\dis_{\delta\gamma\beta}$ , 
\\
$H^{\dis\dis}_{4} = \nabla^\alpha\dis^{\beta\gamma\delta} \nabla_\alpha\dis_{\gamma\beta\delta}$ , 
& $H^{\dis\dis}_{5} = \nabla^\alpha\dis^{\beta\gamma\delta} \nabla_\alpha\dis_{\delta\beta\gamma}$ , 
\\
$H^{\dis\dis}_{6} = \nabla^\alpha\trp12^\beta\nabla_\alpha\trp12_\beta$ , 
& $H^{\dis\dis}_{7} = \nabla^\alpha\trp13^\beta\nabla_\alpha\trp13_\beta$ , 
& $H^{\dis\dis}_{8} = \nabla^\alpha\trp23^\beta\nabla_\alpha\trp23_\beta$ , 
\\
$H^{\dis\dis}_{9} = \nabla^\alpha\trp12^\beta\nabla_\alpha\trp13_\beta$ , 
& $H^{\dis\dis}_{10} = \nabla^\alpha\trp12^\beta\nabla_\alpha\trp23_\beta$ , 
& $H^{\dis\dis}_{11} = \nabla^\alpha\trp13^\beta\nabla_\alpha\trp23_\beta$ , 
\\
$H^{\dis\dis}_{12} = \divp1^{\alpha\beta}\divp1_{\alpha\beta}$ , 
& $H^{\dis\dis}_{13} = \divp1^{\alpha\beta}\divp1_{\beta\alpha}$ , 
\\
$H^{\dis\dis}_{14} = \divp2^{\alpha\beta}\divp2_{\alpha\beta}$ , 
& $H^{\dis\dis}_{15} = \divp2^{\alpha\beta}\divp2_{\beta\alpha}$ , 
\\
$H^{\dis\dis}_{16} = \divp3^{\alpha\beta}\divp3_{\alpha\beta}$ , 
& $H^{\dis\dis}_{17} = \divp3^{\alpha\beta}\divp3_{\beta\alpha}$ , 
\\
$H^{\dis\dis}_{18} = \divp1^{\alpha\beta}\divp2_{\alpha\beta}$ , 
& $H^{\dis\dis}_{19} = \divp1^{\alpha\beta}\divp2_{\beta\alpha}$ , 
\\
$H^{\dis\dis}_{20} = \divp1^{\alpha\beta}\divp3_{\alpha\beta}$ , 
& $H^{\dis\dis}_{21} = \divp1^{\alpha\beta}\divp3_{\beta\alpha}$ , 
\\
$H^{\dis\dis}_{22} = \divp2^{\alpha\beta}\divp3_{\alpha\beta}$ , 
& $H^{\dis\dis}_{23} = \divp2^{\alpha\beta}\divp3_{\beta\alpha}$ , 
\\
$H^{\dis\dis}_{24} = \divp1^{\alpha\beta}\nabla_\alpha\trp12_\beta$ , 
& $H^{\dis\dis}_{25} = \divp1^{\alpha\beta}\nabla_\alpha\trp13_\beta$ , 
& $H^{\dis\dis}_{26} = \divp1^{\alpha\beta}\nabla_\alpha\trp23_\beta$ , 
\\
$H^{\dis\dis}_{27} = \divp3^{\alpha\beta}\nabla_\alpha\trp12_\beta$ , 
& $H^{\dis\dis}_{28} = \divp3^{\alpha\beta}\nabla_\alpha\trp13_\beta$ , 
& $H^{\dis\dis}_{29} = \divp3^{\alpha\beta}\nabla_\alpha\trp23_\beta$ , 
\\
$H^{\dis\dis}_{30} = \divp2^{\alpha\beta}\nabla_\beta\trp12_\alpha$ , 
& $H^{\dis\dis}_{31} = \divp2^{\alpha\beta}\nabla_\beta\trp13_\alpha$ , 
& $H^{\dis\dis}_{32} = \divp2^{\alpha\beta}\nabla_\beta\trp23_\alpha$ , 
\\
$H^{\dis\dis}_{33} = (\trdivp1)^2$ , 
& $H^{\dis\dis}_{34} = (\trdivp2)^2$ , 
& $H^{\dis\dis}_{35} = (\trdivp3)^2$ , 
\\
$H^{\dis\dis}_{36} = \trdivp1\, \trdivp2$ , 
& $H^{\dis\dis}_{37} = \trdivp1\, \trdivp3$ , 
& $H^{\dis\dis}_{38} = \trdivp2\, \trdivp3$ \ ;
\end{tabular}
\ee
and 6 independent terms of the $R\nabla\dis$-type:
\be
\begin{tabular}{lll}
$H^{R\dis}_{7} = R^{\alpha\beta}\divp1_{\alpha\beta}$ , 
& $H^{R\dis}_{8} = R^{\alpha\beta}\divp2_{\alpha\beta}$ , 
& $H^{R\dis}_{9} = R^{\alpha\beta}\divp3_{\alpha\beta}$ , 
\\
$H^{R\dis}_{10} =R\, \trdivp1$ , 
& $H^{R\dis}_{11} = R\, \trdivp2$ , 
& $H^{R\dis}_{12} = R\, \trdivp3$ .
\end{tabular}
\label{Rnabphi}
\ee

%%%%%%%%%%%%%%%%%%%%%%%%%%%
\section{\label{sec:app:partial_results}Partial results}

\subsection{The contribution of $\d g$, $\d \theta$ and $\d A$}

Here we present separately the contribution of the first term in \eqref{beta_functional_with_ghosts}, which is somewhat more compact in the Einstein basis.
\begingroup
\small
\allowdisplaybreaks
\begin{align*} 
B_4&(\Delta_{g\theta\! A}) = - \frac{1}{32 \pi^2} \int d^4 x \sqrt{g} 
\left[
\left(\frac{95}{6} + \frac{19 a_0^2}{8 a_1^2} - \frac{13 a_0}{4 a_1}\right) H^{RR}_1 + \left(7 + \frac{11 a_0^2}{6 a_1^2} + \frac{13 a_0}{3 a_1}\right) H^{RR}_2 \right. \\ & 
+ \left(- \frac{37}{12} + \frac{a_0^2}{12 a_1^2} - \frac{5 a_0}{12 a_1}\right) H^{RR}_3 - \frac{11}{3} H^{R\dis}_{7} + \left(\frac{99}{2} + \frac{34 a_0^2}{3 a_1^2} - \frac{61 a_0}{6 a_1}\right) H^{R\dis}_{8} 
\\ & 
+ \left(- \frac{259}{6} - \frac{34 a_0^2}{3 a_1^2} + \frac{73 a_0}{6 a_1}\right) H^{R\dis}_{9} 
+ \left(\frac{1}{3} + \frac{a_0}{3 a_1}\right) H^{R\dis}_{10} + \left(- \frac{47}{12} + \frac{13 a_0^2}{12 a_1^2} + \frac{5 a_0}{12 a_1}\right) H^{R\dis}_{11} \\ & + \left(\frac{29}{12} - \frac{13 a_0^2}{12 a_1^2} - \frac{7 a_0}{4 a_1}\right) H^{R\dis}_{12} + \left(\frac{7}{6} + \frac{17 a_0^2}{24 a_1^2} + \frac{a_0}{2 a_1} - \frac{a_4}{6 a_1}\right) H^{\dis\dis}_{1} \\ & + \left(- \frac{20}{3} - \frac{53 a_0^2}{24 a_1^2} + \frac{5 a_0}{2 a_1} - \frac{a_4}{6 a_1}\right) H^{\dis\dis}_{2} + \left(\frac{37}{24} + \frac{a_0^2}{6 a_1^2} - \frac{13 a_0}{24 a_1}\right) H^{\dis\dis}_{3} 
\label{B_XYZ} \numberthis
\\ & + \left(\frac{43}{24} + \frac{a_0^2}{6 a_1^2} - \frac{15 a_0}{8 a_1}\right) H^{\dis\dis}_{4} + \left(- \frac{9}{4} - \frac{a_0^2}{3 a_1^2} + \frac{29 a_0}{12 a_1}\right) H^{\dis\dis}_{5} + \left(\frac{2}{3} + \frac{5 a_0^2}{12 a_1^2} + \frac{9 a_0}{8 a_1}\right) H^{\dis\dis}_{6} 
\\ & + \left(- \frac{7}{6} + \frac{5 a_0^2}{12 a_1^2} + \frac{a_0}{8 a_1}\right) H^{\dis\dis}_{7} + \frac{1}{2} H^{\dis\dis}_{8} + \left(1 - \frac{5 a_0^2}{6 a_1^2} - \frac{5 a_0}{4 a_1}\right) H^{\dis\dis}_{9} + \left(- \frac{11}{6} - \frac{a_0}{a_1}\right) H^{\dis\dis}_{10} \\ & + \left(\frac{5}{6} + \frac{a_0}{a_1}\right) H^{\dis\dis}_{11} + \left(- \frac{4}{3} - \frac{17 a_0^2}{24 a_1^2} - \frac{a_0}{2 a_1} + \frac{2 a_4}{3 a_1}\right) H^{\dis\dis}_{12} + \left(\frac{55}{6} + \frac{53 a_0^2}{24 a_1^2} - \frac{5 a_0}{2 a_1} + \frac{2 a_4}{3 a_1}\right) H^{\dis\dis}_{13} \\ & + \left(\frac{15}{8} + \frac{7 a_0^2}{12 a_1^2} - \frac{3 a_0}{4 a_1}\right) H^{\dis\dis}_{14} + \left(\frac{5}{24} + \frac{5 a_0^2}{8 a_1^2} - \frac{5 a_0}{8 a_1}\right) H^{\dis\dis}_{15} + \left(- \frac{13}{8} + \frac{7 a_0^2}{12 a_1^2} - \frac{7 a_0}{12 a_1}\right) H^{\dis\dis}_{16} \\ & + \left(\frac{29}{24} + \frac{5 a_0^2}{8 a_1^2} - \frac{11 a_0}{24 a_1}\right) H^{\dis\dis}_{17} + \left(- \frac{19}{4} - \frac{a_0^2}{3 a_1^2} + \frac{53 a_0}{12 a_1}\right) H^{\dis\dis}_{18} + \left(\frac{25}{12} + \frac{a_0^2}{3 a_1^2} - \frac{37 a_0}{12 a_1}\right) H^{\dis\dis}_{19} \\ & + \left(\frac{47}{12} + \frac{a_0^2}{3 a_1^2} - \frac{37 a_0}{12 a_1}\right) H^{\dis\dis}_{20} + \left(- \frac{23}{12} - \frac{a_0^2}{3 a_1^2} + \frac{7 a_0}{4 a_1}\right) H^{\dis\dis}_{21} + \left(\frac{5}{12} - \frac{7 a_0^2}{6 a_1^2} + \frac{4 a_0}{3 a_1}\right) H^{\dis\dis}_{22} \\ & + \left(- \frac{7}{12} - \frac{5 a_0^2}{4 a_1^2} + \frac{13 a_0}{12 a_1}\right) H^{\dis\dis}_{23} + \left(\frac{3}{2} - \frac{5 a_0^2}{6 a_1^2} - \frac{35 a_0}{12 a_1}\right) H^{\dis\dis}_{24} + \left(- \frac{13}{6} + \frac{5 a_0^2}{6 a_1^2} + \frac{23 a_0}{12 a_1}\right) H^{\dis\dis}_{25} \\ & + \left(\frac{1}{6} + \frac{a_0}{a_1}\right) H^{\dis\dis}_{26} + \left(-3 + \frac{5 a_0^2}{6 a_1^2} + \frac{7 a_0}{4 a_1}\right) H^{\dis\dis}_{27} + \left(\frac{13}{3} - \frac{5 a_0^2}{6 a_1^2} - \frac{3 a_0}{4 a_1}\right) H^{\dis\dis}_{28} \\ & + \left(- \frac{5}{6} - \frac{a_0}{a_1}\right) H^{\dis\dis}_{29} + \left(\frac{1}{6} + \frac{a_0}{6 a_1}\right) H^{\dis\dis}_{30} + \left(- \frac{5}{6} - \frac{a_0}{6 a_1}\right) H^{\dis\dis}_{31} + \frac{2}{3} H^{\dis\dis}_{32} + \frac{5}{6} H^{\dis\dis}_{33} \\ & + \left(\frac{13}{24} + \frac{a_0^2}{8 a_1^2} - \frac{3 a_0}{8 a_1}\right) H^{\dis\dis}_{34} + \left(\frac{29}{24} + \frac{a_0^2}{8 a_1^2} - \frac{a_0}{24 a_1}\right) H^{\dis\dis}_{35} + \left(\frac{1}{3} - \frac{2 a_0}{3 a_1}\right) H^{\dis\dis}_{36} 
\\ & \left.
+ \left(-1 + \frac{2 a_0}{3 a_1}\right) H^{\dis\dis}_{37} + \left(- \frac{17}{12} - \frac{a_0^2}{4 a_1^2} + \frac{5 a_0}{12 a_1}\right) H^{\dis\dis}_{38} 
+ \dots \right] \, .
\end{align*}
\endgroup

%%%%%%%%%%%%%%%%%%%%%%%%%%%%%%%%%%%%%%%%%%%
\subsection{The contribution of the ghosts}
%%%%%%%%%%%%%%%%%%%%%%%%%%%%%%%%%%%%%%%%%%%

The ghost operator can be written in the form
\be
-\begin{pmatrix}
1& 0\\
0 & 1
\end{pmatrix}
\nabla^2
+
\begin{pmatrix}
V_{LL}^\m& V_{LD}^\m\\
V_{DL}^\m & V_{DD}^\m
\end{pmatrix}
\nabla_\m
+
\begin{pmatrix}
W_{LL}^\m& W_{LD}^\m\\
W_{DL}^\m & W_{DD}^\m
\end{pmatrix}\ ,
\ee
where the coefficients $V$ and $W$ can be read off from (\ref{ghost_op}).
We observe the special property
$$
\nabla_\m V_{AB}^\m=W_{AB} \, ,
$$
for $A,B=L,D$.
Its contribution to the logarithmically divergent part of the effective action can be written in the Einstein form as
{\small
\bear \label{b4DeltaGhostEinsteinRPhi}
B_4 (\Delta_{gh}) &= - \frac{1}{32 \pi^2} \int d^4 x \sqrt{g} \left[
- \frac{923}{1440} H^{RR}_1 - \frac{101}{720} H^{RR}_2 + \frac{1}{9} H^{RR}_3 + \frac{1}{6} H^{R\dis}_{7} 
- \frac{35}{24} H^{R\dis}_{8} 
\right. \\ &
+ \frac{13}{12} H^{R\dis}_{9} 
- \frac{7}{48} H^{R\dis}_{11} + \frac{5}{24} H^{R\dis}_{12} 
+ \frac{1}{3} H^{\dis\dis}_{2} - \frac{1}{48} H^{\dis\dis}_{4} - \frac{1}{6} H^{\dis\dis}_{5} - \frac{1}{12} H^{\dis\dis}_{8} \\ & + \frac{1}{6} H^{\dis\dis}_{9} + \frac{2}{3} H^{\dis\dis}_{13} + \frac{1}{48} H^{\dis\dis}_{14} + \frac{1}{48} H^{\dis\dis}_{15} + \frac{5}{24} H^{\dis\dis}_{17} + \frac{1}{24} H^{\dis\dis}_{18} 
+ \frac{1}{4} H^{\dis\dis}_{19} 
\\ & \left.
+ \frac{1}{6} H^{\dis\dis}_{20} + \frac{1}{24} H^{\dis\dis}_{23} - \frac{1}{4} H^{\dis\dis}_{25} - \frac{1}{6} H^{\dis\dis}_{27} - \frac{5}{24} H^{\dis\dis}_{31} - \frac{1}{6} H^{\dis\dis}_{33} + \frac{1}{48} H^{\dis\dis}_{34}
+ \dots \right] \, .
\eear
}
The same expression in the Cartan form is
{\small
\bear \label{b4DeltaGhostCartan}
B_4(\Delta_{gh}) &= - \frac{1}{32 \pi^2} \int d^4 x \sqrt{g} \left[
- \frac{3}{4} L^{FF}_{1} 
+ \frac{101}{120} L^{FF}_{2} 
- \frac{1897}{720} L^{FF}_{3} 
+ \frac{31}{15} L^{FF}_{4} 
- \frac{469}{120} L^{FF}_{5} 
\right. \\ & 
+ \frac{287}{240} L^{FF}_{6} 
- \frac{13}{15} L^{FF}_{7} 
+ \frac{247}{180} L^{FF}_{8} 
- \frac{287}{240} L^{FF}_{9} 
+ \frac{403}{720} L^{FF}_{10} 
- \frac{81}{80} L^{FF}_{11} 
+ \frac{46}{45} L^{FF}_{12} 
\\ &
+ \frac{47}{120} L^{FF}_{13}
+ \frac{47}{20} L^{FF}_{14} 
+ \frac{43}{60} L^{FF}_{15} 
+ \frac{1}{9} L^{FF}_{16} 
+ \frac{3}{2} L^{FT}_{1} 
+ \frac{1}{60} L^{FT}_{10} 
+ \frac{49}{30} L^{FT}_{11} 
+ \frac{26}{15} L^{FT}_{12} 
\\ & 
+ \frac{139}{30} L^{FT}_{13}
- \frac{16}{5} L^{FT}_{14} 
+ \frac{1}{15} L^{FT}_{15} 
- \frac{29}{10} L^{FT}_{17} 
- \frac{89}{60} L^{FT}_{21} 
- \frac{287}{240} L^{FQ}_{10} 
+ \frac{13}{20} L^{FQ}_{11} 
\\ & 
+ \frac{33}{40} L^{FQ}_{12} 
+ \frac{47}{20} L^{FQ}_{14} 
- \frac{11}{24} L^{FQ}_{16} 
+ \frac{107}{120} L^{FQ}_{17} 
+ \frac{413}{240} L^{FQ}_{18} 
- \frac{97}{60} L^{FQ}_{19} 
+ \frac{89}{120} L^{FQ}_{23} 
\\ &
- \frac{59}{160} L^{TT}_{1} 
- \frac{167}{80} L^{TT}_{2} 
+ \frac{13}{15} L^{TT}_{3} 
+ \frac{47}{20} L^{TT}_{5} 
+ \frac{13}{30} L^{TQ}_{1} 
+ \frac{47}{30} L^{TQ}_{10}
- \frac{47}{20} L^{TQ}_{11} 
\\ & \left.
- \frac{13}{15} L^{TQ}_{12} 
+ \frac{73}{480} L^{QQ}_{1} 
- \frac{73}{240} L^{QQ}_{10} 
- \frac{47}{120} L^{QQ}_{11} 
+ \frac{73}{120} L^{QQ}_{12} 
+ \frac{37}{240} L^{QQ}_{14}
+ \dots \right] \, .
\eear
}

\section{\label{sec:app:eoms_Donoghue}On-shell reduction equations}

Since in this note we are interested in the behavior of the propagator, we will focus on reducing the number of independent contributions to it.
In the following, we will write down equations neglecting any contributions to the potential and terms of mass dimension higher than 4.
We will denote the corresponding ``equality at the level of the flat space propagator" as $\simeq$.

\begingroup
\small
\allowdisplaybreaks
\begin{align*} 
& 
L^{FF}_{1} - 2 L^{FF}_{3} + 2 L^{FF}_{6} - 2 L^{FF}_{9} - 2 L^{FF}_{12} + 2 L^{FF}_{15} + 2 L^{FT}_{1} + 2 L^{FT}_{10} - 2 L^{FT}_{11} \\ & + 2 L^{FT}_{13} + 2 L^{FT}_{14} - 2 L^{FT}_{15} - 2 L^{FT}_{17} - 2 L^{FQ}_{10} - 2 L^{FQ}_{16} + 2 L^{FQ}_{17} + 2 L^{FQ}_{18} \simeq 0 \, ,
\\ & 
L^{FF}_{1} + 2 L^{FF}_{2} + 2 L^{FF}_{3} - 2 L^{FF}_{6} + 2 L^{FF}_{9} + 2 L^{FF}_{12} - 2 L^{FF}_{15} - 2 L^{FT}_{1} - 2 L^{FT}_{10} + 2 L^{FT}_{11} \\ & - 2 L^{FT}_{13} - 2 L^{FT}_{14} + 2 L^{FT}_{15} + 2 L^{FT}_{17} + 2 L^{FQ}_{10} + 2 L^{FQ}_{16} - 2 L^{FQ}_{17} - 2 L^{FQ}_{18} \simeq 0 \, , 
\\ & 
2 L^{FF}_{13} \simeq 0 \, , 
\\ & 
- L^{FF}_{1} - 2 L^{FF}_{3} + 4 L^{FF}_{4} + 2 L^{FF}_{6} - 2 L^{FF}_{9} - 2 L^{FF}_{12} + 2 L^{FF}_{15} + 2 L^{FT}_{1} + 2 L^{FT}_{10} - 2 L^{FT}_{11} \\ & + 2 L^{FT}_{13} - 2 L^{FT}_{14} - 2 L^{FT}_{15} - 2 L^{FT}_{17} - 2 L^{FQ}_{10} + 2 L^{FQ}_{16} + 2 L^{FQ}_{17} - 2 L^{FQ}_{18} \simeq 0 \, , 
\\ & 
L^{FF}_{1} + 2 L^{FF}_{3} + 4 L^{FF}_{5} - 2 L^{FF}_{6} + 2 L^{FF}_{9} + 2 L^{FF}_{12} - 2 L^{FF}_{15} - 2 L^{FT}_{1} - 2 L^{FT}_{10} + 2 L^{FT}_{11} \\ & - 2 L^{FT}_{13} + 2 L^{FT}_{14} + 2 L^{FT}_{15} + 2 L^{FT}_{17} + 2 L^{FQ}_{10} + 2 L^{FQ}_{16} - 2 L^{FQ}_{17} - 2 L^{FQ}_{18} \simeq 0 \, , 
\label{on-shell_reduction_equations} \numberthis \\ & 
-4 L^{FF}_{14} + 4 L^{FF}_{15} - 2 L^{FQ}_{12} - 2 L^{FQ}_{23} \simeq 0 \, , 
\\ & 
-2 L^{FQ}_{12} \simeq 0 \, , 
\\ \nonumber & 
- L^{FF}_{1} + 2 L^{FF}_{3} + 4 L^{FF}_{5} + 2 L^{FF}_{6} - 2 L^{FF}_{9} - 4 L^{FF}_{11} + 2 L^{FF}_{12} - 2 L^{FF}_{15} + 2 L^{FT}_{1} - 2 L^{FT}_{10} \\ & + 2 L^{FT}_{11} - 2 L^{FT}_{13} - 2 L^{FT}_{14} + 2 L^{FT}_{15} + 2 L^{FT}_{17} - 2 L^{FQ}_{10} + 2 L^{FQ}_{16} + 2 L^{FQ}_{17} - 2 L^{FQ}_{18} \simeq 0 \, , 
\\ & 
L^{FF}_{1} - 2 L^{FF}_{3} + 2 L^{FF}_{6} - 2 L^{FF}_{9} - 2 L^{FF}_{12} + 2 L^{FF}_{15} - 2 L^{FT}_{1} + 2 L^{FT}_{10} - 2 L^{FT}_{11} + 2 L^{FT}_{13} \\ & + 2 L^{FT}_{14} - 2 L^{FT}_{15} - 2 L^{FT}_{17} - 2 L^{FQ}_{10} - 2 L^{FQ}_{16} + 2 L^{FQ}_{17} + 2 L^{FQ}_{18} \simeq 0 \, , 
\\ & 
-2 L^{FF}_{13} - 4 L^{FT}_{12} + 4 L^{FQ}_{14} \simeq 0 \, , 
\\ & 
4 L^{FF}_{8} + 4 L^{FF}_{11} - 4 L^{FF}_{15} + 4 L^{FT}_{11} + 4 L^{FT}_{17} - 4 L^{FT}_{21} + 2 L^{FQ}_{12} - 4 L^{FQ}_{18} + 2 L^{FQ}_{23} \simeq 0 \, , 
\\ &
4 L^{FT}_{10} - 4 L^{FQ}_{10} + 2 L^{FQ}_{12} \simeq 0 \, , 
\\ &
2 L^{FF}_{13} + 4 L^{FT}_{12} \simeq 0 \, , 
\\ & 
4 L^{FF}_{7} - 4 L^{FF}_{8} + 4 L^{FF}_{15} - 4 L^{FT}_{11} - 4 L^{FT}_{17} + 4 L^{FT}_{21} - 2 L^{FQ}_{12} - 2 L^{FQ}_{23} \simeq 0 \, , 
\\ &
-4 L^{FT}_{10} - 2 L^{FQ}_{12} \simeq 0 \, .
\end{align*}
\endgroup

\end{appendix}
\end{CJK*} %Chinese characters

\makeatletter  % forbids page breaks in bibliography
\interlinepenalty=10000  % forbids page breaks in bibliography
\bibliographystyle{jhep-mymod}
\bibliography{main}
\makeatother  % forbids page breaks in bibliography
%\printbibliography

\end{document}